\documentclass[onecolumn]{article}
\usepackage[a4paper, total={6in, 8in}]{geometry}
\usepackage[round]{natbib}
\usepackage[affil-it]{authblk}

\usepackage{graphicx}
\usepackage{svg}
\usepackage{algorithmic}
\usepackage{textcomp}
\usepackage{xcolor}
\usepackage[utf8]{inputenc}
\usepackage{booktabs}       
\usepackage{amsmath,amssymb,amsfonts,mathrsfs}

\usepackage{nicefrac}       
\usepackage{microtype}      
\usepackage{lipsum}
\usepackage{multirow}
\usepackage{enumitem}
\usepackage{placeins}
\usepackage{comment}
\usepackage{pdfpages}

\usepackage[font=footnotesize]{caption}
\usepackage{color}
\definecolor{myblue}{rgb}{0, 0, 0.7}
\usepackage{hyperref}
\usepackage{nameref}
\hypersetup{
    colorlinks,
    citecolor=myblue,
    filecolor=myblue,
    linkcolor=myblue,
    urlcolor=myblue
}

\newcommand{\softmax}{\mathbf{s}}

\begin{document}
\title{PaccMann\textsuperscript{RL}: Designing anticancer drugs from transcriptomic data via reinforcement learning}
%
%
\newcommand\note[1]{\textcolor{black}{#1}}
\author[1,2]{Jannis Born\textsuperscript{*}}
\author[1]{Matteo Manica\textsuperscript{*}}
\author[1]{Ali Oskooei\textsuperscript{*}}
\author[1]{Joris Cadow}
\author[2]{Karsten Borgwardt}
\author[1]{María Rodríguez Martínez}
\affil[1]{IBM Research Zurich, Switzerland}
\affil[2]{Machine Learning \& Computational Biology Lab, ETH Zurich, Switzerland}
%
%

%

\maketitle              
\textsuperscript{*} Equal contributions
\smallskip
\newline

\note{\textbf{125-character summary:} A framework to bridge systems biology and drug discovery for designing anticancer drugs from transcriptomic data}

\begin{abstract} 
With the advent of deep generative models in computational chemistry, in silico drug design has undergone an unprecedented transformation. While state-of-the-art deep learning approaches have shown potential in generating compounds with desired chemical properties, they disregard the cellular environment and biomolecular properties of the target disease. 
Here, we introduce a novel 
framework 
for de-novo molecular design
that systematically leverages systems biology information into the drug discovery process. 
Embodied through two separate Variational Autoencoders (VAE), the drug generation is driven through a disease context (transcriptomic profiles of cancer cells) deemed to represent the target environment in which the drug has to act.
Showcased at the challenging task of de-novo anticancer drug discovery, our conditional generative model is demonstrated to be capable of tailoring anticancer compounds to target a specific biomolecular profile, according to the critic.
Without incorporating explicit information about anticancer drugs, we demonstrate how the molecule generation, starting from a random point in a chemical space, can be biased towards compounds with high predicted inhibitory effect against individual cell lines or cell lines from specific cancer sites.
We verify our approach by investigating candidate drugs generated against specific cancer types and find the highest structural similarity to existing compounds with known efficacy against these cancer types.
Despite no direct optimization of other pharmacological properties, we report good agreement with known cancer drugs in metrics like drug-likeness, synthesizability and solubility.
We envision our approach to be a step towards
increasing success rates in lead compound discovery
and finding more targeted medicines by leveraging the cellular environment of the disease.
\end{abstract}

\section{Introduction}
\label{sec:intro}

\begin{figure}[!htb]
\centering
    \includegraphics[width=1.0\textwidth]{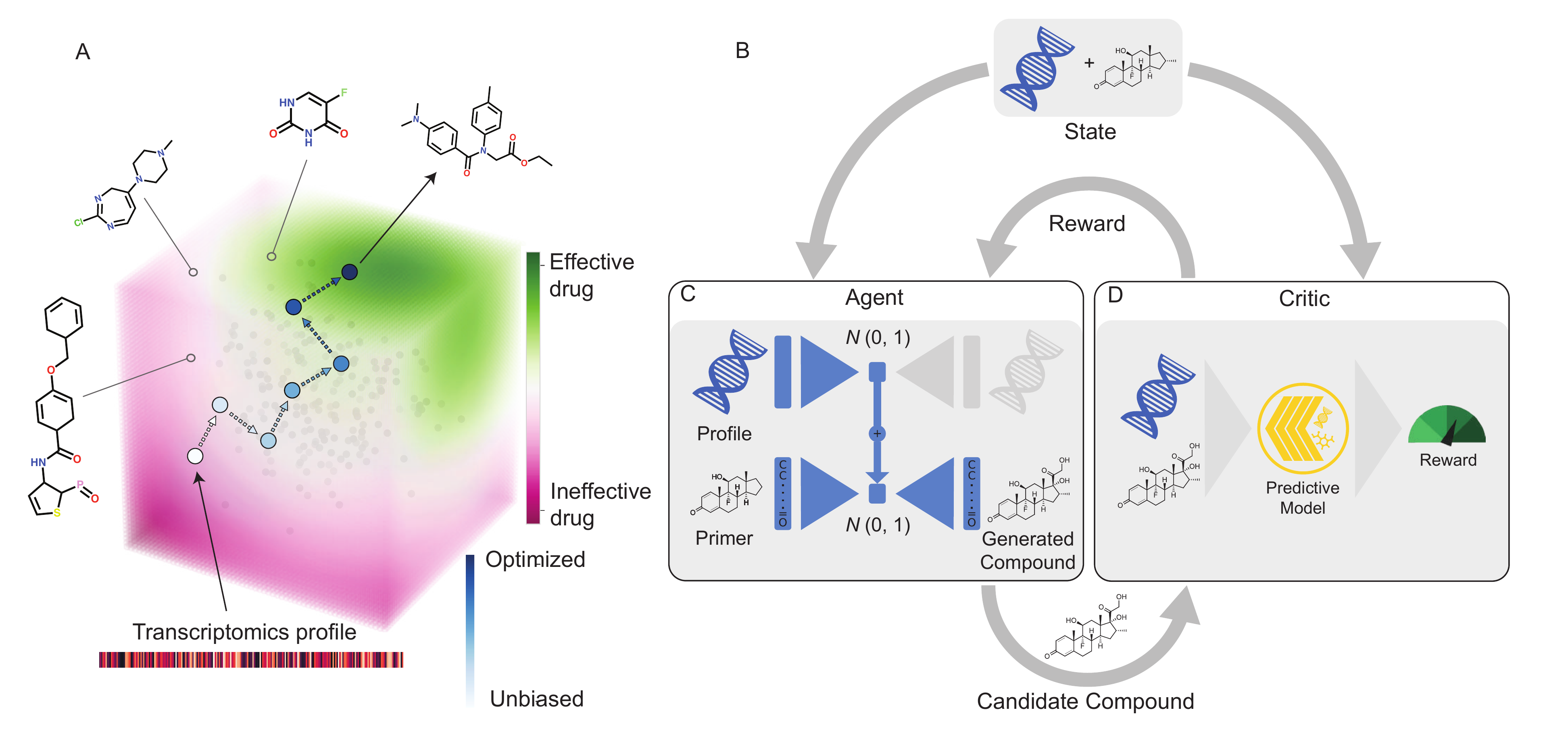}\\
    \caption{
    \textbf{The proposed framework for anticancer compound design against specific cancer profiles.\\}
    \textbf{A)} Conceptually, the model performs a guided walk through the chemical space in order to find effective compounds.
    Starting from an unbiased molecule generator (trained only on a dataset of bioactive compounds without any information about cancer), compounds are sampled and screened in-silico against the transcriptomic profiles of interest. 
    The outcome of the screening guides the generator towards sampling from manifolds with more effective compounds. 
    \textbf{B)}
    The training process depicted in more detail.
    The conditional compound generator (called "agent") is embodied through two, initially separate VAEs. 
    The compound generation starts with a biomolecular profile of interest e.g. a transcriptomic profile.
    Through a pretrained omics VAE, the profile is encoded into the latent space of gene expression profiles.
    The latent representation of the profile is decoded through the molecular decoder of a separately pretrained molecule VAE to produce a candidate compound (see \textbf{C}).
    This generative process can optionally be ``primed'' through encoding a known, effective compound or a functional group with the molecular encoder.
    The proposed compound is then evaluated by a critic, where our critic is represented by a multimodal drug sensitivity prediction model that ingests the compound and the target profile of interest (see \textbf{D}).
    The IC50 efficacy as predicted by the critic, is interpreted as reward and is subject to maximization during the RL based optimization.
    Over the course of training, the generator will thus learn to produce candidate compounds with higher and higher efficacy.
    \texttt{<START>} is the start and \texttt{<END>} is the end token.
    }
    \label{fig:overview}
\end{figure}

Eroom's Law describes the observation that the productivity of the drug discovery pipeline, as measured by the number of FDA approved drugs per invested billion US dollar, halves every nine years since the 1950s~\citep{scannell2012diagnosing}.
Indeed, only a minimal portion of synthesized drug candidates obtain market approval (less than 0.01\%), with an estimated 10-15 years until market release and costs that range between one~\citep{scannell2012diagnosing} to three billion dollars per drug~\citep{schneider2019mind}.
This low efficiency has been attributed to the high dropout rate of candidate molecules in the early stages of the pipeline, highlighting the need for more accurate in silico and in vitro models that produce more potent candidate drugs.
In addition to the initial wet-lab validations, the discovery pipeline involves a sequential process that builds upon high-throughput screenings, ADMET-assessments
and a lengthy phase of clinical trials.
The costs of the experimental and clinical phase can be prohibitive and any solution that helps to reduce the number of required experimental assays can provide a competitive advantage and reduce time to market.
The problem's linchpin is on how to
improve the exploration and navigation through the chemical space that has been estimated to contain $\sim 10^{30}$-$10^{60}$ drug-like molecules with bioactive properties~\citep{polishchuk2013estimation}.
For this task, deep learning methods have recently gained popularity~\citep{chen2018rise} and many have demonstrated the feasibility of in silico design of novel candidate compounds with desired chemical properties~\citep{popova2018deep,gomez2018,you2018graph}.
In all of these models, the generative process is controlled via a structurally driven evaluator (or critic) that biases the generation of a chemical to satisfy the required chemical structural properties.
While very effective in generating compounds with desired chemical properties, these methods disregard system-level information, e.g. about the cellular environment in which the drug is intended to act.
However, the two main causes of the increasing attrition rate in drug design are lacking efficacy against the specific disease of interest and off-target cytotoxicity~\citep{Wehling2009}, calling to bridge systems biology closer with drug discovery. 
To this end, we present a novel framework for anticancer molecule generation based on deep generative models and reinforcement learning that, for the first time, enables generation of candidate compounds while taking into account the disease context encoded in the form of gene expression profile (GEP) of the tumor cell (for a graphical illustration see ~\autoref{fig:overview}A). 

Related methodology has been used for protein-targeting de-novo generation~\citep{zhavoronkov2019deep,aumentado2018latent,grechishnikova2019transformer}. 
These contributions attempt to utilize deep learning methods for de-novo design of compounds to specifically target a protein that has been implicated in tumor proliferation or treatment response (e.g. gene-knockout study).
For example, the study by \citet{zhavoronkov2019deep} curated and utilized, amongst others, patent data and several datasets about molecules (unspecific bioactive compounds, kinase inhibitors, DDR1 kinase inhibitors, molecules targeting non-kinase targets) specifically to develop DDR1 inhibitors.
They synthesised and tested six drug candidates in cell assays, where two were found to be active and one was even successfully validated in animal models.
Envisioning a precision or even personalized medicine perspective, identifying protein targets is challenging, whereas sequencing and omics data are straightforward to gather. 

Very recently, \citet{mendez2020novo} proposed a method for de-novo design of molecules against desired targets, represented by the gene expression signatures of knocked-out (suspected) targets.
However, 97\% of anticancer candidate drugs fail in clinical trials and never receive FDA approval, questioning the current approaches of target identification for the discovery of pharmaceuticals~\citep{wong2019estimation}.
Taking as many as 10 drug-indication-pairs from ongoing clinical trials, \citet{lin2019off} found that the proposed mechanism of action (MOA) of \textit{all} of them were incorrect; knocking out the target genes did not ever hamper cancer fitness.
While the wrong target genes were identified through RNA interference with siRNA seemingly silencing essential off-target genes, all drugs retained their anticancer effect through target-independent mechanisms.
Off-target cytotoxicity being a common MOA of anticancer drugs in clinical trials corroborates to scrutinize current lead compound discovery strategies and calls to develop novel methodology with unconventional approaches.
It is for this reason that we herein propose a novel framework to generate lead compound candidates solely based on a tumour's metabolic signature, as opposed to attempting to target a specific protein or incorporating information about potential targets directly into the design process.
Acting as metabolic signature, we instead guide the learning process solely by transcriptomic profiles of cancer cells,
since transcriptomic data has been successfully used for de-novo drug identification~\citep{verbist2015using, de2018high} and has been advocated for a pivotal role for the future of drug discovery~\citep{dopazo2014genomics}.

Our framework is depicted in \autoref{fig:overview}B and consists of a conditional molecule generator (embodied by two separate VAEs) and a critic module that evaluates the efficacy of proposed compounds on the target profile (see \autoref{fig:overview}D).
The training procedure splits into two stages.
In the first stage, the models are trained independently; one VAE is trained on gene expression data from TCGA~\citep{TCGA_2013}, another VAE is trained on bioactive small molecules from ChEMBL~\citep{chembldatabase} (see~\autoref{fig:overview}C).
As critic, a multimodal drug sensitivity prediction model is fetched from previous work~\citep{manica2019paccmann}. 
In the second stage, the encoder of the profile VAE is combined with the decoder of the molecule VAE and exposed to a joint retraining that is optimized in a policy gradient regime with a reward coming from the critic module.
The goal of the optimization is to tune the generative model such that it generates (novel) compounds that have maximal efficacy against a given biomolecular profile that is characteristic for a cancer site, a patient subgroup or even an individual.
By \textit{efficacy}, we refer to predicted cellular IC50 (i.e. the micromolar concentration necessary to inhibit 50\% of the cells) as opposed to e.g. enzymatic IC50. 
Importantly, this efficacy is a joint property of a drug-cell-pair; as treatment response to a compound heavily varies depending on the tumor's genomic and transcriptomic makeup~\citep{geeleher2016cancer}.
In this work, we emphasize profile-specific compound generation and optimize the generator with IC50 as sole critic.

Here, we present a first step towards our vision of a generic framework where the molecule generation can be conditioned on possibly multimodal context information such as a (multi)omics profiles, a primed drug or a drug scaffold, a target protein or any combination thereof.
The resulting latent spaces, consisting of semantically distinct ontological entities, could then be jointly explored by machine learning techniques designed to operate on sets instead of fixed-length vectors, such as permutation-invariant operations~\citep{zaheer2017deep}. \\

\FloatBarrier

\section{Results}
\label{sec:results}

\paragraph{Pretraining Profile VAE and SMILES VAE.}
In the first phase of training, the two components of \autoref{fig:overview}C  were trained independently. 
The profile VAE (PVAE) consisted of a set of stacked dense layers and was trained as a denoising VAE to enhance generalization abilities.
The purpose of the PVAE was to find a lower dimensional representation of the cell profiles that maintains structural similarity and later allows a fusion with the latent representation of molecules.
The encoder of the PVAE learned to embed gene expression profiles (bulk RNA-Seq from TCGA~\citep{TCGA_2013}) meaningfully into a latent space, such that the decoder could reconstruct the profiles, but also generate novel, seemingly realistic gene expression profiles (GEP). 

The SMILES VAE (SVAE) was pretrained for 10 epochs on $\sim$1.4 million structures from ChEMBL~\citep{chembldatabase}.
Both encoder and decoder consisted of stack-augmented gated-recurrent units as used in~\citep{popova2018deep}.
The purpose of the SVAE was to learn the syntax of SMILES and general semantics about bioactive compounds.
The novelty and diversity of the generated molecules was validated by sampling 10,000 molecules through decoding random points from the latent space. 
96.2\% of the 10,000 generated molecules were valid molecular structures (surpassing the results of~\citet{popova2018deep} who used the same stack-augmented GRUs  and reported 95\% SMILES validity) and 99.72\% of the valid molecules were unique across the 10,000 generations.
Comparing the Tanimoto similarity (i.e. the jaccard index, a well-established chemical structure similarity measure~\citet{Tanimoto1958elementary}) of the molecular fingerprints (ECFP~\citet{rogers_extended-connectivity_2010}) of 1000 generated molecules with the training and test data from ChEMBL, we find that the vast majority had a Tanimoto similarity ($\tau$) between 0.2 and 0.6 (on average $0.41\pm0.1$ for training and $0.38\pm0.08$ for testing molecules) suggesting that our model learned to propose novel molecular structures from the chemical space of about 10\textsuperscript{30} to 10\textsuperscript{60} molecules~\citep{polishchuk2013estimation}.
In addition, a \href{https://paccmann.github.io/rl/unbiased.html}{visualization of the chemical space} of ChEMBL as well as generated compounds through the TMAP algorithm (a library that visualizes high dimensional data through minimum spanning trees~\citep{probst2019visualization}) showed that the generated molecules mix well with the training molecules into the chemical space. 

For detailed results of both the PVAE and the SVAE, please see the~\href{ssec:resultspretraining}{appendix (S5)}.

\vspace{-2mm}
\subsection{Disease-specific compound generation}
\label{ssec:paccmann_rl_results_generator}
Herein, we present the results of our drug generator conditioned on gene expression profiles of cancer subtypes.
\begin{figure*}[!htb]
    \centering
    \includegraphics[width=.9\textwidth]{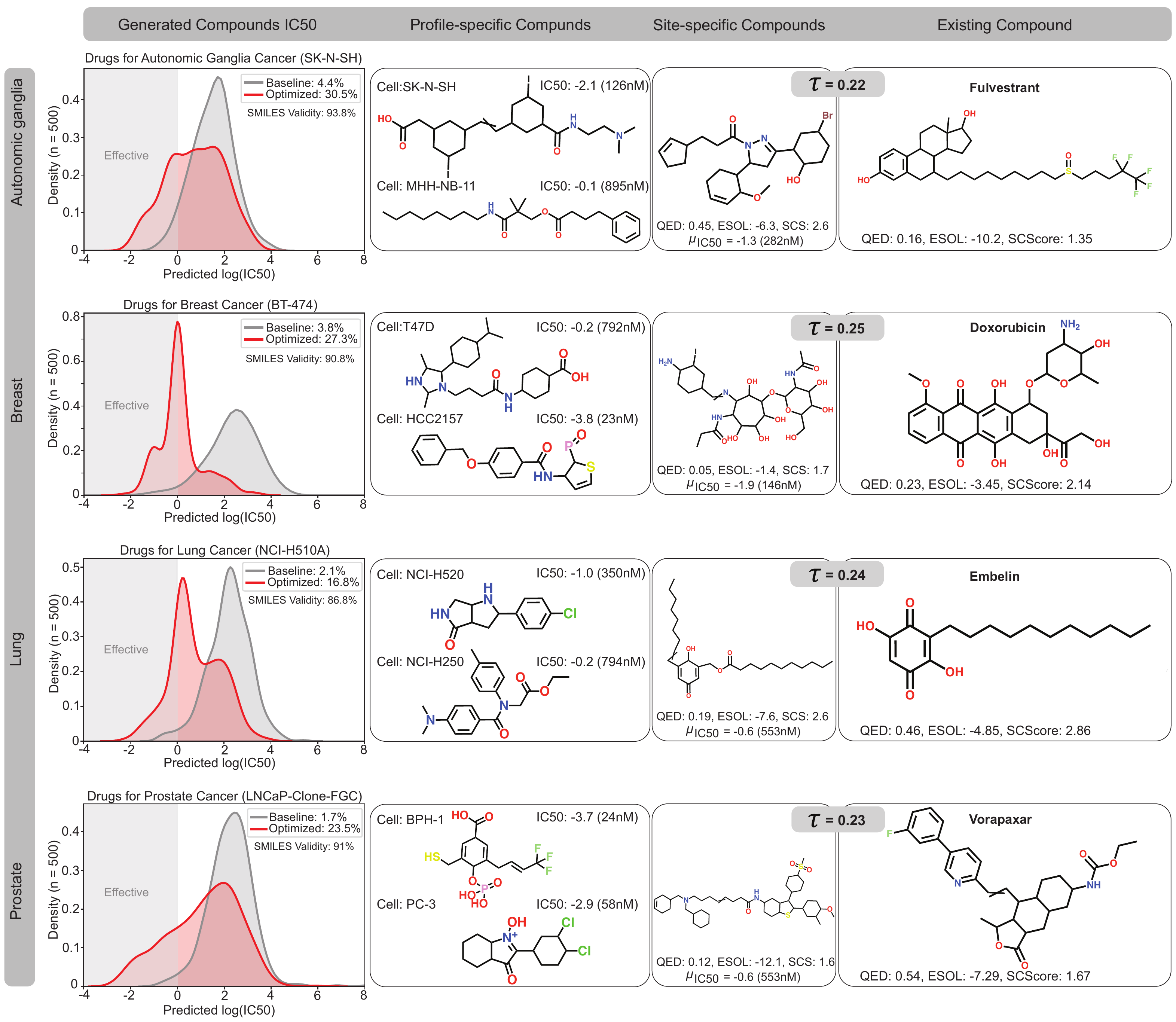}\\
    \caption{
    \textbf{Sample results for profile-driven model optimization and anticancer compound generation.}
    Each row illustrates the results of training the RL pipeline on cell lines from a specific cancer type: autonomic ganglia, breast, lung and prostate cancer.
    The first column compares the distributions of IC50 predictions given by the critic model for a set of $n$=500 drug candidates generated with RL optimization and without RL optimization.
    As demonstrated by the density plots, the RL optimization process leads to candidate compounds with a lower mean IC50 for the target cancer -- highlighting the successful optimization of the generative model towards the design of more effective compounds.
    The second column presents candidate compounds with a high predicted efficacy (low IC50) against a particular cell line that was not seen during training (this corresponds to a "personalized medicine" regime)
    The third column showcases generated compounds that were optimized to be effective against all cell-line profiles of the given cancer type in each row (corresponds to a "precision medicine" regime).
    In the fourth column, we present an \textit{existing} anticancer compound (approved against at least one type of cancer), that was in the top-3 neighborhood of the generated compound in the third column.
    The existing and generated compounds are compared in terms of Tanimoto structural similarity as well as three chemical scores crucial in drug design namely, druglikeness (QED, 0 worst, 1 best), synthetic complexity (SCS or SCScore, 1 best, 5 worst) and solubility (ESOL, given in $M/L$).
    }
    \label{fig:rl_generator_panel}
\end{figure*}
As a proof of concept, we show results for cancer in four different sites: breast (carcinoma), lung (carcinoma), prostate (carcinoma) and autonomic ganglia (neuroblastoma).
The conditional generator was initialized as the SVAE, i.e. sampling from the unbiased generator yielded random molecules from the chemical space as learned from the ChEMBL data.
For the evaluation, all generated compounds with a predicted IC50 value below 1$\mu$M were considered to be \textit{effective}.
Moreover, within each cancer type (or site), 80\% of the cell lines (breast: 50, lung: 169, prostate: 7, autonomic ganglia: 56) were considered as training cell lines and used to optimize the parameters $\Theta$ of the conditional generator.
We observed that over time the generator learned to produce more drugs with higher predicted efficacy according to the critic.
To test both the generalization abilities and whether the generator actually utilized the omics-profile for the generation, we used the remaining 20\% of cell lines to verify whether conditioning the generator on unseen cell lines of the same site also leads to compounds with low IC50.
As presented in \autoref{fig:rl_generator_panel} (left column), our model learned to produce compounds with lower IC50 values, for unseen cell lines from the given cancer site.
In other words, the IC50 distribution of candidate compounds proposed by the generative model were successfully shifted towards higher efficacy (lower IC50).
The baseline model corresponds to the pretrained SVAE from which $n=500$ molecules were randomly sampled.
In all four cases, a significant portion (between 17\% and 30\%) of molecules generated from the optimized model were assigned a IC50 value below $1\mu$M, whereas only 1-4\% of the candidates generated by the baseline model (i.e. the SVAE) were classified as effective.
Moreover, in all cases the generator maintained almost an equal SMILES validity (87\%-94\%) compared to the baseline, much higher than what \citet{mendez2020novo} reported based on gene expression (8-9\%).
The second column of \autoref{fig:rl_generator_panel} shows generated molecules that are predicted as being effective against unseen cell lines from the respective cancer site.
As opposed to the personalized regime in the second column, the third column of \autoref{fig:rl_generator_panel} showcases a precision medicine regime. 
Here, novel molecules were designed specifically for each cancer site i.e. a single, characteristic GEP.
In all cases, the model generated compounds that exhibited high efficacy against the average cellular profile of the target site while maintaining efficacy against the majority of individual cell lines for that site.
\paragraph{Investigation of nearest neighbors}
For a more quantitative assessment, the last column of \autoref{fig:rl_generator_panel} compares the four cancer type-specific candidate compounds with one of their top-3 neighbors using the Tanimoto similarity score, $\tau$, from several hundreds of existing anticancer compounds. 
It is well known that Tanimoto similarity across compounds is highly correlated with their induced sensitivity patterns on cancer cell lines~\citep{shivakumar2009structural}.
The candidate compound proposed against breast cancer (\autoref{fig:rl_generator_panel} second row, third column) has Doxorubicin, a commonly used chemotherapeutical against breast cancer~\citep{lao2013liposomal}, as one of the top-3 nearest neighbors.
The generated compound against lung cancer (\autoref{fig:rl_generator_panel}, third row, third column) presents similarities to Embelin, an existing anticancer compound from the GDSC database.
Comparing the two structures, it is evident that the generated compound and Embelin share a long carbon chain and a single six-membered fully carbonic ring.
Embelin was tested against 965 cell lines from GDSC/CCLE from which the highest reported efficacy is against a lung cell line (NT2-D1).
Embelin is also known to be the only known non-peptide inhibitor of XIAP~\citep{poojari2014embelin}, a protein that plays an important role in lung cancer development~\citep{cheng2010xiap}.
The closest neighbor of the prostate-specific generated compound (\autoref{fig:rl_generator_panel} fourth row, third column) is Vorapaxar. Its efficacy is highest against a prostate cancer cell line (DU\_145) according to GDSC/CCLE.
Vorapaxar is an antagonist of a protease-activated receptor (PAR-1) that is known to be overexpressed in various types of cancer, including prostate~\citep{zhang2009protease}.
Lastly, the third closest neighbor of the generated compound against neuroblastoma (\autoref{fig:rl_generator_panel} first row, third column) is Fulvestrant, an antagonist/modulator of ER$\alpha$ which has recently been proposed as a novel anticancer agent for neuroblastoma~\citep{gorska2016nitro}.
To summarise, for all four investigated cancer types, the proposed compounds some high structural similarity to anticancer drugs that are, for each specific cancer type, either 1) already FDA approved (breast), 2) known inhibitors of relevant targets (lung, prostate) or 3) have been advocated for (neuroblastoma). 
This result is remarkable, specially  as the generator was never exposed to any anticancer compounds. Indeed only the critic had seen two out of the four compounds during training, highlighting the fact that the generator has \emph{de novo} learn the structural characteristics that make a compound efficacious against a particular cancer type.

In the above literature review, the search space was restricted to compounds with known anticancer properties.
To investigate whether the proposed compounds had generally a higher similarity to drugs associated with cancer, we carried out a comparison with compounds from a broader pool of chemicals, namely ChEMBL~\citep{chembldatabase}, a database of $>1.5$ million bioactive molecules with drug-like properties. 
The nearest neighbour ($\tau=0.54$) of our breast compound in the ChEMBL database is \href{https://www.ebi.ac.uk/chembl/compound_report_card/CHEMBL1093122/}{CHEMBL1093122}, a conjugate of plumbagin and phenyl-2-amino-1-thioglucoside that inhibits the synthesis of mycothiol~\citep{gammon2010conjugates}.
Plumbagin itself and many of its derivatives are widely studied anti breast cancer compounds~\citep{zhang2016plumbagin,kawiak2017plumbagin,dandawate2014anticancer}.
For our lung cancer compound, the nearest neighbor ($\tau=0.48$) is \href{https://pubchem.ncbi.nlm.nih.gov/compound/5378708}{polyoxyethylene dioleate}, a surfactant that has been patented for the treatment of eight types of cancer including three types of lung cancer (small lung cell cancer, lung adenocarcinoma and metastatic lung cancer, ~\citet{girsh2007lipid}). 
It is also utilised in targeted drug delivery systems for drug-resistant lung cancer~\citep{kaur2016surfactant}.
The nearest neighbour ($\tau=0.31$) of the prostate cancer compound is \href{https://www.ebi.ac.uk/chembl/compound_report_card/CHEMBL2104127/}{Clinolamide} (or Linolexamide) which is included in a patent of diagnostic and/or therapeutically active compounds for several types of cancer, including prostate cancer~\citep{klaveness2004diagnostic}.
The nearest neighbor ($\tau=0.35$) of the proposed neuroblastoma compound is \href{https://pubchem.ncbi.nlm.nih.gov/compound/401736}{NSC-715466}.
NSC-715466 has been evaluated for anticancer effects in the NCI-60 database~\citep{shoemaker2006nci60} and inhibits cancer cell growth by $65\%\pm 15\%$ across all tested cell lines, with a below-averge inhibition for cancer in the central nervous system ($57\%\pm9\%$). 
Regarding its efficacy, it only falls in the 51st percentile of all 53,217 compounds tested in NCI-60, which presumably prevented further investigations.
The four discussed ChEMBL compounds as well as the analysis of NSC-715466 can be found in the \hyperref[ssec:nnchembl]{appendix (S6)}.
It is promising to observe that  the molecules with the highest Tanimoto similarity to our compounds are associated with cancer (some even to the specific types of cancer our compounds were optimized for) even using a larger database of bioactive compounds. 
However, it is worth keeping in mind that a high Tanimoto score to a known cancer drug is not necessary for anticancer drug efficacy, as some  cancer drugs used for the same cancer type or even sharing the same mechanism of action exhibit low Tanimoto similarity, e.g. TKIs used for NSCLC such as Crizotinib and Erlotinib, $\tau=0.11$.
Across all anticancer compounds in GDSC and CCLE databases, we note that the average Tanimoto similarity ($\tau=0.149\pm0.05$) is not much lower than the average similarity of two compounds of a given site ($\tau=0.154\pm0.06$). 
For the following results and discussion,  the anticancer compounds from GDSC/CCLE were associated with the site where they had the highest average IC50 efficacy. 

\begin{figure*}[!htb]
    \includegraphics[width=1.\textwidth]{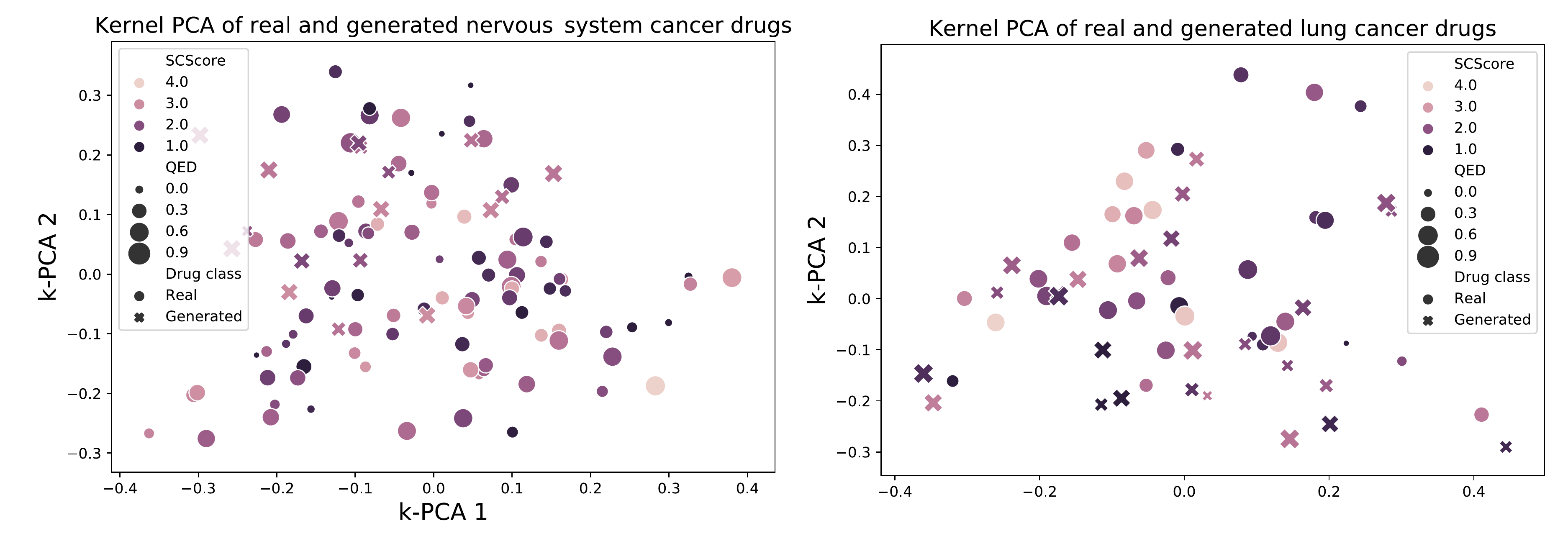}\\
    \vspace{-5mm}
    \caption{
    \textbf{Visualization of generated and real anticancer drugs.}
    A kernel PCA of real and generated molecules based on Tanimoto similarity. 
    The size of the points is denoted the QED score while the coloring represents the synthetic complexity score (SCScore).
    Overall, both generated and existing molecules are heterogeneously distributed in the 2D projection and do not form clear clusters.
    }
    \label{fig:visualize_drugs}
\end{figure*}

To understand better whether the generated drugs mimic the space of cancer-specific anticancer drugs, \autoref{fig:visualize_drugs} shows visualizations of real and generated cancer drugs for one specific cancer type, using kernel PCA based on Tanimoto similarity~\citep{scholkopf1998nonlinear}.
In addition to the class belongings, the plots also depict the QED score~\citep{bickerton2012quantifying}, a quantitative estimate of drug-likeness (0 worst, 1 best), and SCScore~\citep{coley2018scscore}, an estimated score of synthetic complexity (1 best, 5 worst).
The fact that the generated molecules are well intermingled with the real drugs suggests that the generator proposes diversified structures that mimic some properties of anticancer compounds.
It is also curios to see that several real drugs have low QED and/or high synthetic complexity scores (the same holds for the generated molecules).
Moreover, we provide interactive TMAP visualizations of the site-specific, generated compounds (links in \href{sec:availability}{availability section}).

\paragraph{Chemical properties of generated molecules}
\label{ssec:chemicalscores}
In this work, the conditional generator is trained using PaccMann as sole critic.
However, besides inhibitory efficacy, there is a myriad of properties of a candidate drug that crucially influence its potential for becoming an anticancer compound.

\begin{figure*}
    \centering
    \includegraphics[width=1\textwidth]{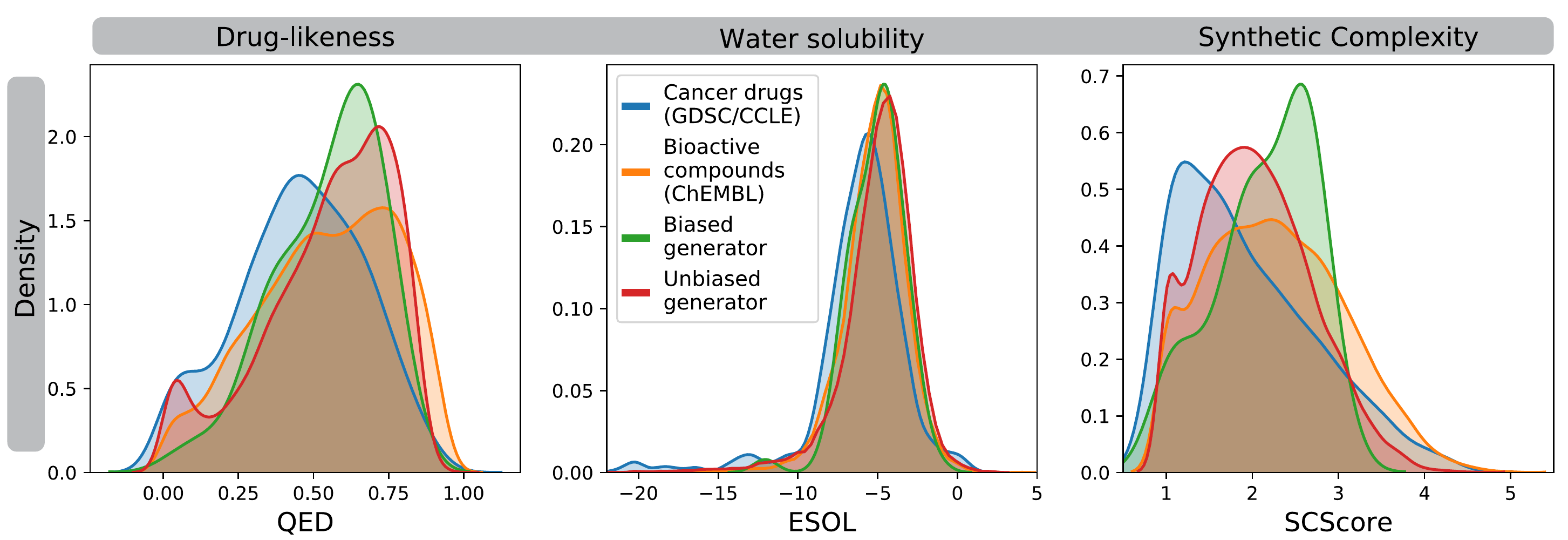}\\
    \caption{
    Comparison of chemical scores for real drugs in GDSC and CCLE database versus our generated compounds.
    We compared three chemical scores for druglikeness as assessed by QED score (0 worst, 1 best), for solubility as assessed via ESOL, given in $\log(M/L)$ (most drugs have a solubility between -8 and -2) and for synthetic accessibility as assessed by SAS (1 best, 10 worst).
    These three scores are computed for the panel of known anticancer drugs, bioactive molecules from ChEMBL, molecules generated before (red) and after (green) RL optimization. }
    \label{fig:rl_generator_scores}
\end{figure*}

Some of these can be approximated in-silico, e.g. water solubility (ESOL, ~\citet{delaney2004esol}), drug-likeness (QED,~\citet{bickerton2012quantifying}) and synthesizability (SCScore,~\citet{coley2018scscore}).
\autoref{fig:rl_generator_scores} gives an overview about the distribution of QED, ESOL and SCScore for sets of 1) known anticancer compounds (blue), 2) molecules from ChEMBL (orange), 3) compounds generated by the SVAE (red) and 4) compounds proposed by the conditional generator (green).
Despite none of these properties was explicitly optimized, comparing the distributions reveals overall a good agreement.
Interestingly, anticancer drugs exhibit, compared to the ChEMBL compounds, on average much less drug-like properties (lower QED) and seem easier to synthesize (lower SCScore).
This tendency of anticancer drugs for synthetically less complex structures is likely to result from the high attrition rate in clinical trials and the corresponding cost reduction policies.
It is also encouraging to see that the unbiased generator (SVAE) generates on average molecules with more desired properties compared to the data used for training (ChEMBL compounds have on average lower QED and higher SCScore). \\
Moreover, the cancer drugs have, on average a significantly lower QED than the other three sets ($0.45\pm0.2$ with $>10\%$ of GDSC/CCLE drugs even having a QED $<0.2$ whereas it is $0.55$ for the other three sets).
Indeed the QED scores of the other three sets were so similar that we failed to reject the null hypothesis that the QED scores of these three sets are from different distributions (Kruskal-Wallis test, $\alpha=0.05$).
Regarding synthetic complexity, both the unbiased and the biased generator fail to produce molecules with SCScores as low as the anticancer drugs (MWU, $p<0.01$), but they produce structures that are estimated to be less complex than the ChEMBL molecules (MWU, $p<0.01$).
Overall, it can be seen that the biased generator produces molecules with less desired properties than the unbiased generator (SVAE).
This is expected because the unbiased generator was pretrained to mimic the data from ChEMBL whereas no explicit optimization of chemical scores was performed during the RL optimization.
For one generated compound, we exemplary show a possible synthesis route that was assigned a high confidence score by the retrosynthesis model~\citep{ibmrxn} and consists of four reactions and 10 commercially available reactants~(see \hyperref[ssec:retrosynthesis]{appendix (S8)}).

\citet{savjani2012drug} reported that 40\% of novel chemicals cause practical problems due to insolubility. 
Water solubility remains challenging to approximate in-silico~\citep{sorkun2019aqsoldb} and thus we treat with caution the good agreement in the ESOL scores~\autoref{fig:rl_generator_scores} (middle panel).

Finally, we would like to point out that utilizing a IC50 drug sensitivity prediction model as sole critic limits the performance  of the entire pipeline, as 
the expressive power of the conditional generator is inherently upper bounded by the predictive power of the critic.
Crucially, due to a lack of available data, the critic was only trained on anticancer drugs but not on compounds without inhibitory efficacy against cell lines.
We therefore verified the generalization capabilities of the critic by comparing the predicted efficacy of cancer drugs (most of them were seen during training) and a "negative" set of molecules from ChEMBL across all 965 cell lines from GDSC.
The results can be found in \hyperref[ssec:paccmannchembl]{appendix S7} and show that while 15.2\% of the virtual drug screenings with anticancer compounds were positive (IC50$<1\mu$mol), only 2.2\% of ChEMBL molecules showed potential anticancer effects.
Moreover, the generated anticancer compounds were found to have a significantly higher Tanimoto similarity to anticancer drugs than to both ChEMBL molecules ($p<0.01$, one-sided MWU) and molecules generated without the RL optimization, i.e. from the SVAE ($p<0.01$, one-sided MWU).
These two results are encouraging. They suggest that PaccMann can seemingly drive the molecule generation away from ordinary, bioactive compounds as in ChEMBL more towards mimicking the properties of actual anticancer drugs.

\section{Discussion}
\label{sec:discussion}

We herein presented the first framework for anti-cancer compound generator that enables us to condition the molecular generation on the biomolecular profile (specifically we explored transcriptomic profiles) of the target cell or cancer site.
We demonstrated, using a RL optimization framework, that our proposed generative model could be optimized to produce candidate compounds with high predicted efficacy (IC50) against a given target profile, even if this profile was never seen during training.
Notably, this was achieved despite the fact that the generator was never exposed to anticancer drugs explicitly, but only pretrained on bioactive compounds from ChEMBL. 
The only component that has been trained on drugs with known anticancer effects is the critic, which only communicates with the generator by providing a reward function. 

An analysis of the generated compounds for four different cancer types demonstrated that  the predicted compounds  share many structural similarities with known anticancer compounds for the same cancer types that the generated compound was optimized for.

While our results are a promising stepping stone for profile-specific anticancer compound generation, further optimization must be done before it can be used a reliable tool for drug discovery.
For instance, other properties of a candidate drug other than inhibitory efficacy that determine its potential for becoming a successful anticancer compound, for example water solubility, drug-likeness, synthesizability, environmental toxicity or off-target cytotoxicity are not directly optimized.
However, despite not explicitly incorporating them into the reward function, we find that the produced molecules exhibit desired properties in terms of drug-likeness, water solubility and ease of synthesis.

Furthermore,  the high attrition rate in drug discovery has been  attributed to either a lack of efficacy or off-target cytotoxicity~\citep{Wehling2009}, the latter implying that very often the mechanism of action of an active compound has been incorrectly characterized. 
In that respect, the design of new AI-enhanced drug design approaches that can bypass the need of a detailed characterization of drug targets and cytotoxicity mechanisms can greatly improve current drug discovery pipelines. Future work should focus on incorporating information in the reward function not only about drug efficacy but also about other drug-relevant chemical properties and predicted off-target cytotoxicity effects.
The resulting multimodal objectives may be difficult to optimize due to possibly counteracting/interfering gradients. A possible approach to  circumvent this challenge is by using explicit compensation techniques~\citep{yu2020gradient} or defining gradient-free global objectives~\citep{hase2018chimera}. 
Another challenge that needs to be overcome to improve  the reliability and accuracy of the critic is the expected distributional differences between the data used for training  (cancer cell lines) and the targeted data  (human data from clinical trials). A possible approach that can be explored is the exploitation of  transfer learning techniques, as suggested in~\citet{SharifiNoghabi2020}. 

Oftentimes, medical chemists do not start the drug design from scratch, but from the scaffold of an approved drug.
The goal of the scaffold hopping is to find a drug with similar effects (e.g. increased efficacy or reduced side effects). 
While our framework enables users to incorporate  prior knowledge into the design process by priming the latent code, we have not yet explored the full potential of this idea herein. As the decoded molecule is not guaranteed to maintain similarity to the primer, the recently proposed "deep generative scaffold decorator" could be integrated into our framework to facilitate a more systematic exploration and the possibility of adding fragments to established drug scaffolds~\citep{arus2020smiles}.
An alternative future endeavor is to explore graph-based instead of sequential representations of molecules  to directly generate a molecular graph from the context set using a conditional structure generation framework~\citep{yang2019conditional}.


\footnotesize{
\section{Methods}
\label{sec:methods}


\begin{figure}[!htb]
\centering
    \includegraphics[width=0.8\textwidth]{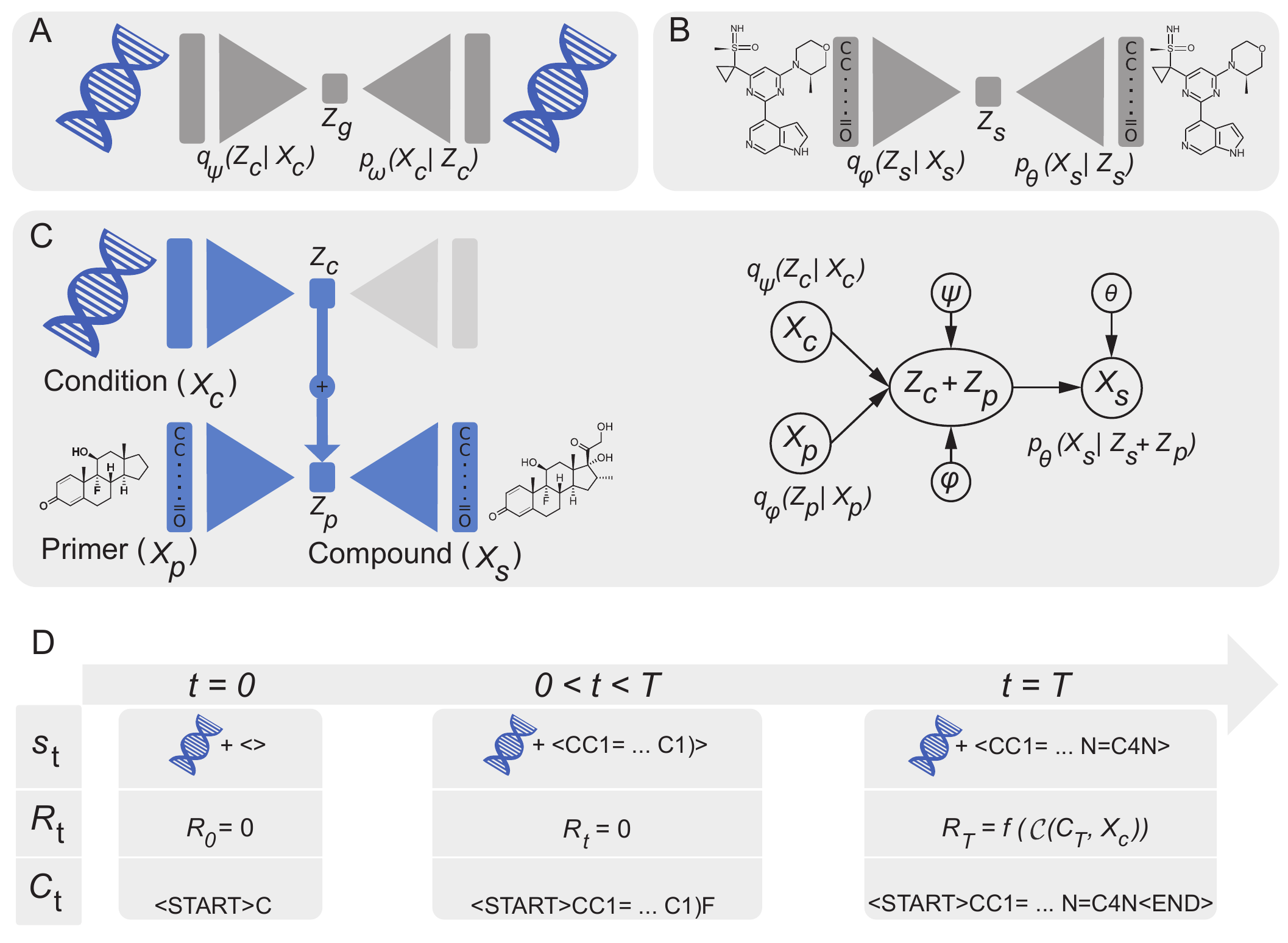}\\
    \caption{
    \textbf{Architectural details of the conditional drug generator.} \\
    \textbf{A)} A biomolecular profile VAE (PVAE) was pretrained on RNA-Seq data from TCGA to encode a transcriptomic profile $X_c$ into a latent code $Z_c$, before attempting to decode $X_c$ from it. 
    \textbf{B)} Similarily, a sequential compound generator VAE (SVAE) was trained to encode and decode SMILES representations $X_s$ of molecules.
    \textbf{C)} PVAE and SVAE are combined to obtain a conditional molecule generator.
    As shown in the graphical model, the combination is achieved by using a permutation-invariant operation (e.g. addition) to fuse the latent spaces of omics profiles and molecules to a joint, multimodal representation.
    \textbf{D)} Molecules are generated directly as SMILES sequences and are assembled in a sequential process, one token at a time.
    A full cycle of this process includes a state ($s_{t}$, where $s_0=X_c$, i.e. a TCGA RNA-Seq transcriptomic profile), a reward ($R_{t}$) and a generated candidate compound ($C_{t}$).
    }
    \label{fig:architecture}
\end{figure}

\paragraph{Conditional generator ($\mathcal{G}$).}
Our conditional generator is a molecule generator that produces a candidate drug structure represented as its SMILES sequences~\citep{weininger1988smiles}.
SMILES sequences are preferable over (functional) fingerprint-based representations of molecules (e.g. ECFP~\citep{rogers_extended-connectivity_2010}) since they have shown to be superior in both predictive~\citep{jastrzkebski2016learning,manica2019paccmann} and generative models for molecules~\citep{bjerrum2018improving}.
In our use case, the generative process is conditioned on a target biomolecular profile, e.g. from  a patient or a disease.
Inspired by~\citet{gomez2018}, we concluded that variational autoencoders (VAE) are the ideal generative model for our task since by design they bring about a structurally ordered latent space that simplifies the combination of different information sources.
Our conditional generator combines two VAEs that are trained independently prior to being fused together: 1) a denoising VAE for cancer profile encoding/generation (called PVAE, \autoref{fig:architecture}A) and 2) a sequential VAE (SVAE) for SMILES sequence generation (\autoref{fig:architecture}B).
The mathematical formulation of VAEs can be found in~\citet{kingma2013auto,sohn2015learningCVAE}.
PVAE is pretrained on gene expression profiles (GEP) to learn a consistent latent representation for biomolecular signatures.
SVAE is pretrained on bioactive drug-like molecules to learn the syntax of valid SMILES and general molecular semantics.
Generative models of SMILES sequences necessitate the ability to \textit{count}
Models that process SMILES sequences greatly benefit from the ability to \textit{count} the ring opening and closing symbols in a molecule, as a single mistake in the sequential generation of a SMILES renders the entire string invalid.
To circumvent that standard recurrent and convolutional networks lack the proficiency to count, we utilize a stack memory~\citep{Hopcroft1969}, in our case implemented through stack-augmented GRUs as proposed by~\citet{joulin2015inferring} (for equations and other details of the SVAE and the stack see the \hyperref[ssec:svae]{appendix (S1)}).
Thereafter, the encoder of the PVAE is fused with the decoder of the SVAE via their latent space (\autoref{fig:architecture}C).
The combination of the two models enables to learn a latent space that links biomolecular profiles and chemical structures providing an effective way to sample novel compounds given a specific GEP.
In the RL optimization phase, the weights of the fused model (which were pretrained independently) are fine-tuned using a reward from the critic.

\paragraph{Critic ($\mathcal{C}$).}
The critic is a multimodal drug sensitivity prediction model that evaluates the efficacy of any given candidate compound against a biomolecular profile of interest, e.g. gene expression of a cancer cell line. 
The critic outputs a non-negative reward that depends on the candidate compound predicted IC50 for the target profile, such that low IC50 values associated with higher compound efficacy receive higher rewards than high IC50 values. The reward is then used in a RL framework to update the conditional generator.
Following the most recent advances for multimodal drug sensitivity prediction we herein utilize \textit{PaccMann} as a critic~\citet{manica2019paccmann}.

\paragraph{The RL framework.}
The conditional generator is retrained in combination with the critic in a RL-based optimization process to tailor molecules towards the given GEP.
First, the GEP is encoded into a latent space, $Z_{c}$ (see \autoref{fig:architecture}C).
This embedding is then added to the latent encoding of a primer compound or substructure ($Z_{p}$).
The advantage of using a primer is that it enables injection of prior knowledge into the model by starting the generative process from an existing and proven effective compound or functional group -- instead of designing a compound from scratch.
However, this priming is optional and we do not only sample closely around existing compounds but we instead sample a larger fraction of the chemical space.
As can be seen in the graphical model in~\autoref{fig:architecture}C the molecule generation is conditioned on a context $\mathcal{Z}$, where in this work $\mathcal{Z}=\{Z_c, Z_p\}$. 
$Z_c$ and $Z_p$ reflect embeddings learned from semantically different data modalities (gene expression and molecules).
To combine these (latent) representations, we use summation because it is a permutation invariant operation and has been proposed to combine a variable set of unstructured latent encodings~\citep{zaheer2017deep}.
Alternatives include mixup functions such as weighted sums or dimension-wise sampling from a categorical (Bernoulli) distribution~\citep{beckham2019adversarial}.
Our additive latent representation is similar in concept to the conditional VAE with additive Gaussian encoding space~\citep{wang2017diverseCVAE}.
Intuitively, this fusion presumably warps the latent space from encoding structural similarity (of molecules or GEP) into functional similarity so as to aggregate molecules with similar predicted efficacy for a given cell line~\citep{gomez2018}.
Note that using a primer compound or substructure is optional and if no priming compound is used, simply the latent space representation of the \texttt{<START>} token is added to the latent encoding of the target GEP. \\
Next, the conditional generator decodes the latent encoding, $Z_{c}$ + $Z_{p}$, and generates a molecular structure that, in combination with the GEP, is fed to the critic to produce a certain reward for the generated compound, as illustrated in \autoref{fig:architecture}C.
Following the notation of~\citet{popova2018deep}, the conditional generator, $\mathcal{G}$, acts as the \textit{agent} and PaccMann (the multimodal IC50 prediction model, $\mathcal{C}$) represents the \textit{critic}.
The weights of $\mathcal{C}$ are fixed.
We aim to optimize $\Theta$, the parameters of $\mathcal{G}$, to produce candidate compounds, $C_{T}$, that target a specific GEP, $X_{c}$.
In contrast to~\citet{popova2018deep}, we define the set of states $\mathcal{S}$ as all possible SMILES strings (with length $\leq T$) paired with the target GEP.
The set of possible actions $a$ that $\mathcal{G}$ can take is a set $\mathcal{A}$, which is a vocabulary of all characters and symbols of the canonical SMILES language.
As depicted in \autoref{fig:architecture}D, molecules are generated by $\mathcal{G}$ by sampling an action $a_t$ at each step($0<t<T$) from $p(a_t|s_{t-1})$, where $s_{t-1} = (C_{t-1}, X_{c})$ and $C_0$ is simply the \texttt{<START>} token.
Terminal states $S^{*}\subset S$ are reached when either $t=T$ or when the terminal action $a_T =\texttt{<END>}$ has been sampled.
$\mathcal{G}$ is trained to learn a policy, $\Pi(\Theta)$, by maximizing: 
\begin{equation}
    \Pi(\Theta) = \sum_{s_T \in S^*} P_{\Theta}(s_T)R(s_T)
    \label{eq:policy_gradient}
\end{equation}
where $P_{\Theta}(s_T):= \prod_{t=0:T}p(a_t|s_{t-1})$ and the state $s_T = (C_T, X_c)$ is a tuple of the candidate compound $C_T$ and the cell profile $X_c$ and the reward $R(s_T) = f(\mathcal{C}(C_{T}, X_{c}))$ is the output of the critic $C$ scaled by a reward function $f$.
In our experiments, all intermediate rewards $R(s_t)$ are set to 0, since $C_t$ (the intermediate SMILES string) will in almost all cases not resemble a valid molecule.
The sum is approximated using policy gradients, specifically the REINFORCE algorithm~\citep{williams1992simple} and the reward function $f$ for determining the reward from the IC50 prediction, $\mathcal{C}(C_T, X_c)$, is computed by $f(IC50)=\exp{\left(\frac{-IC50}{\alpha}\right)}$ where $\alpha=5$ in this work (see details in the \hyperref[ssec:rewardfunction]{appendix (S2)}).

\paragraph{Data.}
For the PVAE, we employed a training dataset of 11,592 (standardized) RNA-Seq GEPs from healthy and cancerous human tissue from the TCGA database and validated it on 1,289 samples from the same database~\citep{TCGA_2013}.
Since the dataset was too small to train on the full cohort of ~20,000 genes and most genes are correlated to a subset of landmark genes, the number of genes was reduced to the same 2,128 genes as used in~\citet{manica2019paccmann}, following the network propagation procedure described in~\citet{oskooei2019network}.
The SVAE was pretrained on the SMILES representation of 1,576,904 compounds (10\% were held out for performance validation) from the ChEMBL database~\citep{gaulton2016chembl}.
For RL optimization of $\mathcal{G}$, we used GEPs publicly available from GDSC~\citep{gdsc} and CCLE~\citep{ccle} databases.
Since the RNA-Seq of these cancer cell line databases were passed through the PVAE (pretrained on human samples from TCGA~\citep{TCGA_2013}), we compared the standardized gene expression distributions for the selected genes across these databases and found good agreement (see \hyperref[ssec:rnaseqcomparison]{appendix (S3)}), in alignment with the reported consensus between transcriptomic data in CCLE and TCGA~\citep{ghandi2019next}.
To train the critic ($\mathcal{C}$), IC50 drug sensitivity data from GDSC and CCLE was utilized.
The hyperparameter and details on the utilized hardware and software can be found in the  \hyperref[ssec:implementationdetails]{appendix (S4)}.

\section{Availability of software and materials}
\label{sec:availability} 
The omics data used to pretrain the PVAE, the molecular data for the SVAE and the cell profiles used in the RL regime as well as the pretrained models can be found on \href{ https://ibm.box.com/v/paccmann-pytoda-data}{ https://ibm.box.com/v/paccmann-pytoda-data}.
To assess the critic, please see~\citet{manica2019paccmann}.
All code to reproduce the experiments is publicly available on \href{https://github.com/PaccMann/}{https://github.com/PaccMann/}.
For a detailed example see \href{https://github.com/PaccMann/paccmann_rl}{https://github.com/PaccMann/paccmann\_rl}.

The interactive TMAP visualizations of the molecules generated by the (unbiased) SVAE can be found on \href{https://paccmann.github.io/rl/unbiased.html}{https://paccmann.github.io/rl/unbiased.html}.
The TMAPs of the cancer-site-specific candidate compounds are accessible on \href{https://paccmann.github.io/}{https://paccmann.github.io/}.
\section*{Acknowledgements}
The project leading to this publication has received funding from the European Union's Horizon 2020 research and innovation programme under grant agreement No 826121.
J.B. acknowledges financial support from the German Academic Exchange Service (DAAD).

}

\bibliographystyle{plainnat}
\bibliography{main}
\newpage

\newpage
\thispagestyle{plain}
\setcounter{page}{1}
\setcounter{figure}{0}
\setcounter{section}{1}
\renewcommand{\thesubsection}{S\arabic{subsection}}
\renewcommand{\thefigure}{S\arabic{figure}}
\newcommand{\mysection}[2]{\setcounter{subsection}{#1}\addtocounter{subsection}{-1}\subsection{S#2}}

\section*{PaccMann\textsuperscript{RL}: Designing anticancer drugs from transcriptomic data via reinforcement learning - Appendix}

\subsection{SMILES VAE architecture with StackGRU}
\label{ssec:svae}
\begin{figure}[!htb]
    \centering
    \includegraphics[width=1\textwidth]{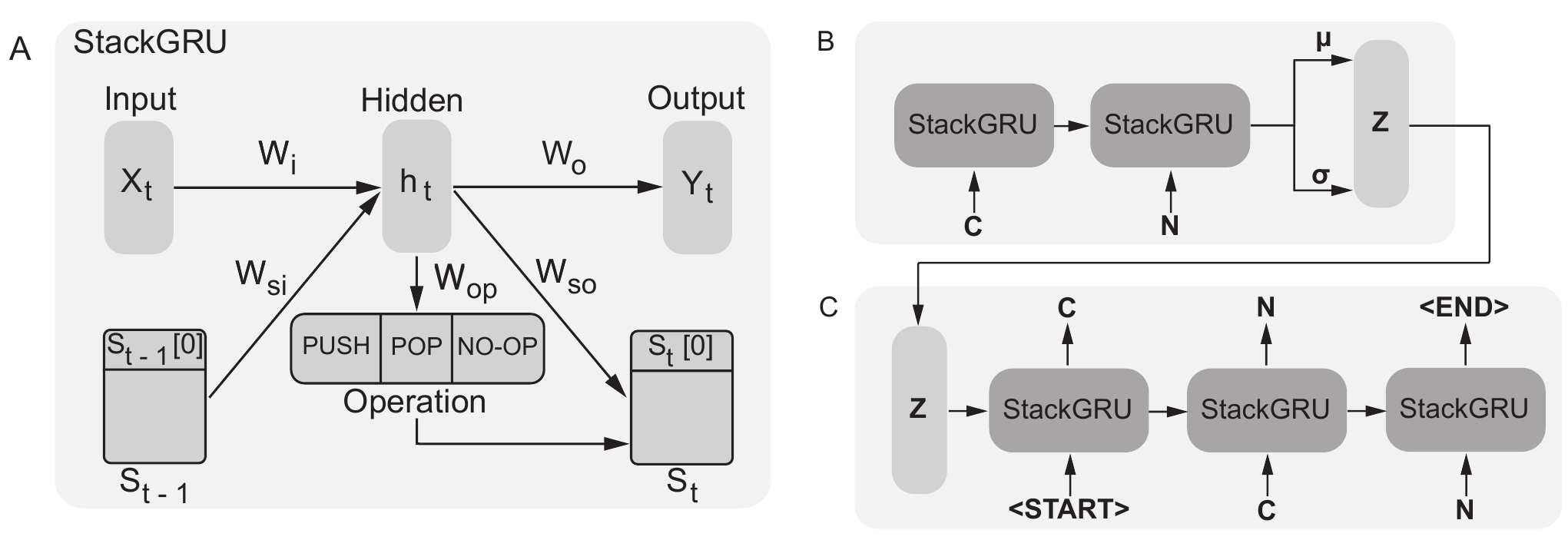}
    \caption{
    (A)
    The StackGRU architecture adopted in the SVAE.
	The stack-augmented GRU (StackGRU) architecture complements a regular GRU with a stack that allows one out of three possible operations at each time-step: \texttt{PUSH}, \texttt{POP} and \texttt{NO-OP}.
    The operation vector is determined through a softmax from the hidden state of each time step.
    \newline
    (B) and (C) are encoder and decoder of the SVAE architecture.
	(B) encodes the SMILES sequences into multivariate Gaussians with parameters $\mu$ and $\sigma$.
	(C) The decoder StackGRU units reconstruct the SMILES sequence from a latent representation ($Z_p$) sampled from the multivariate Gaussian.
    }
    \label{fig:stackgru}
\end{figure}

To enable neural networks to count,~\cite{joulin2015inferring} introduced stack-augmented RNN.
Stack-RNNs complement RNNs with a differentiable push-down stack operated through learnable controllers, $op_{t}$ at step $t$, that involve three operations: \texttt{PUSH}, \texttt{POP} and \texttt{NO-OP} (see \autoref{fig:stackgru}).
\begin{align}
\label{eq:stack_controls}  
    op_{t}\ = \softmax({W}_{op}{h}_{t}),
\end{align}

\noindent
where $h_{t}$ is the hidden state, $W_{op}$ is a $3 \times H$ matrix (H being the dimension of hidden state) and $\softmax$ is the softmax function.
At each time step the controller probabilities are determined from Equation~\ref{eq:stack_controls} and the stack memory is updated using the learned controller via a multiplicative gating mechanism:

\begin{align}
\label{eq:stack_update}  
\small
\begin{cases}
S_{t}[0] &= op_{t}[\texttt{PUSH}]\softmax(W_{so}h_{t}) + op_{t}[\texttt{POP}]S_{t-1}[1] + \\
&\quad op_{t}[\texttt{NO-OP}]S_{t-1}[0]\\
h_{t} &= \softmax(W_{i}X_{t} +W_{R}h_{t-1}+W_{si}S_{t-1})
\end{cases}
\end{align}

\noindent
where $S_t$ is the stack, $W_{so}$ is a $1 \times H$ matrix and $W_{si}$ is a $H \times N$ matrix ($\it{N}$ being the stack height).
$W_i$ is the input matrix applied to the sequence and $W_{R}$ is the recurrent matrix.
It should be noted that for the sake of brevity, we only show the update equation for the topmost element of the stack in Equation~\ref{eq:stack_update}.

\thispagestyle{plain}
\FloatBarrier
\subsection{Reward function}
\label{ssec:rewardfunction}
\begin{figure}[!htb]
    \centering
    \includegraphics[width=0.5\textwidth]{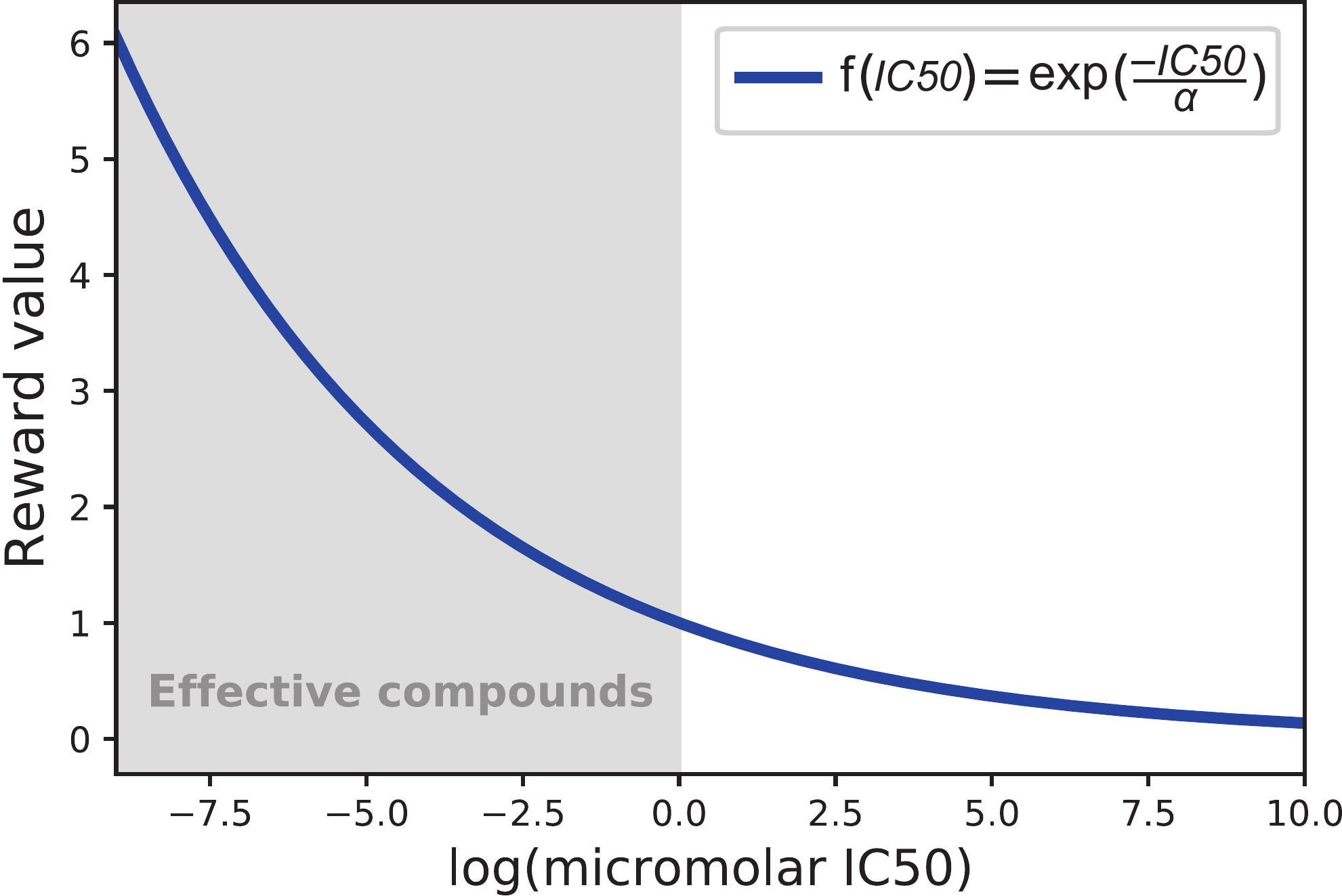}\\
    \caption{
    Reward function to map the predicted IC50 of the critic (PaccMann) to a reward being fed to the conditional generator.
    To produce the plot, $\alpha$ was set to 5.}
    \label{fig:reward_fct}
\end{figure}

A reward function $f$ was used to map the logarithmic micromolar IC50 values predicted by the critic to a reward that was subject to maximization in our adopted RL framework (see \autoref{fig:reward_fct})..
It is computed by $f(IC50)=\exp{\left(\frac{-IC50}{\alpha}\right)}$, where $IC50 = \mathcal{C}(C_T, X_c)$ and $\alpha \in \mathbb{R}^{+}$ is a tunable hyperparameter that determines how much the generator is rewarded for designing effective versus ineffective compounds, i.e., a smaller $\alpha$ leads to a greedier generator.
\FloatBarrier
\subsection{Gene expression in human samples and cancer cell lines}
\label{ssec:rnaseqcomparison}
\begin{figure}[!htb]
    \centering
    \includegraphics[width=0.7\textwidth]{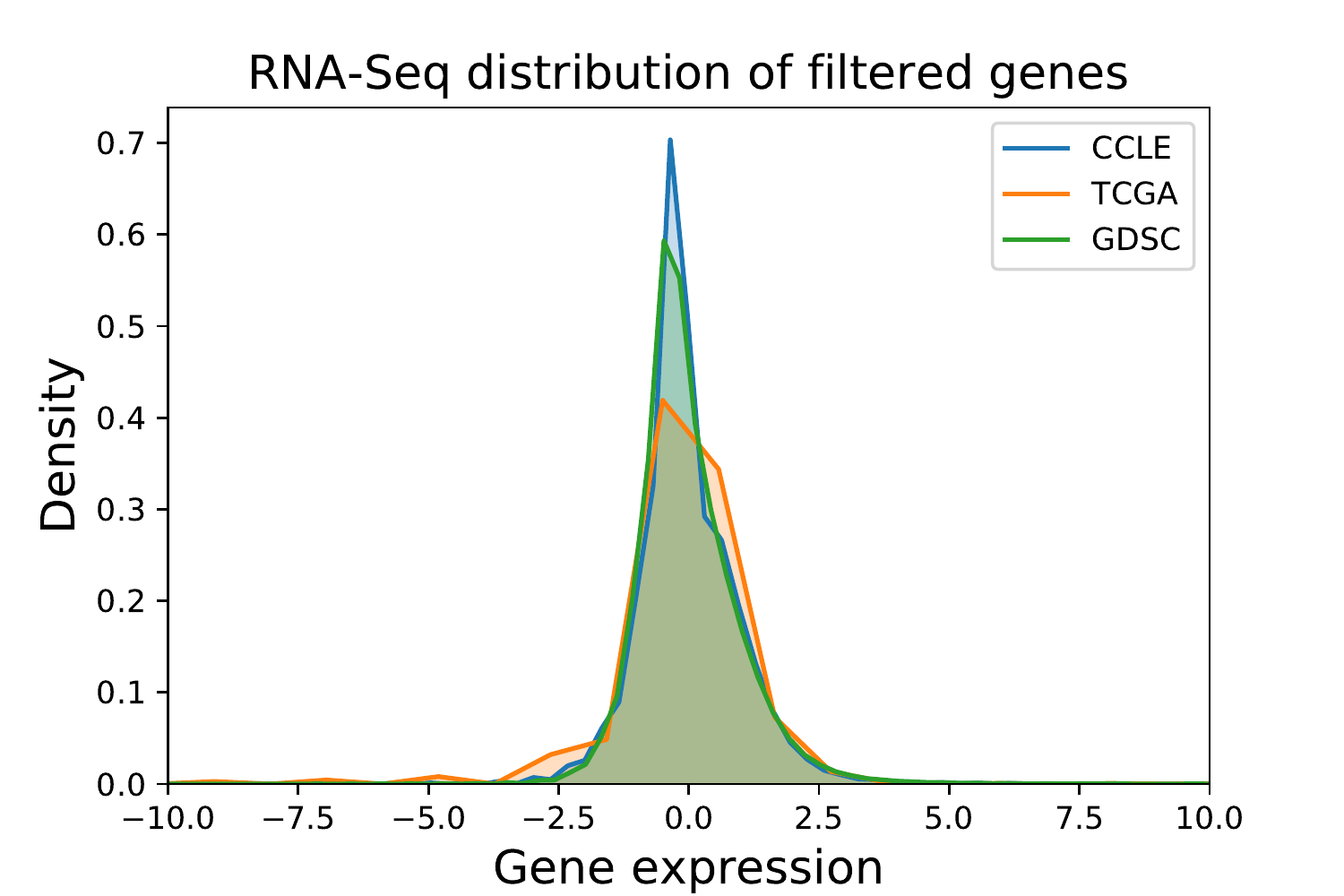}\\
    \caption{
    Distribution of standardized gene expression values across the cancer cell line databases CCLE and GDSC as well as the human sample database TCGA.
}
    \label{fig:rnaseq_comp}
\end{figure}

Comparing the standardized gene expression values of GDSC~\citep{gdsc} and CCLE~\citep{ccle} with the one from human samples from TCGA~\citep{TCGA_2013} reveals a similarity (\autoref{fig:rnaseq_comp}).
This justifies our choice of utilizing the encoder of the PVAE for cell line data during the RL regime, although it was initially pretrained on human samples from TCGA.
\FloatBarrier

\subsection{Implementation and training details}
\label{ssec:implementationdetails}
All models were implemented in \texttt{PyTorch} 1.0 and trained on a cluster equipped with \texttt{POWER8} processors and a \texttt{NVIDIA Tesla P100}. 

\paragraph{PVAE.}
The model consisted of four dense layers of [1024, 512, 256 and 200] units with ReLU activation function and dropout of $p=0.2$ in both, the encoder and the decoder.
The dimensionality of the latent space ($n$) was 128.
We minimized the variational loss, consisting of the reconstruction loss and KL divergence, using Adam optimizer ($\beta_{1}=0.9$, $\beta_{2}=0.999$, $\varepsilon=1\mathrm{e}{-8}$) and a decreasing learning rate starting at $0.001$~\cite{kingma2014adam}.
To further regularize the PVAE, denoising methods were employed by 1) applying a dropout of 0.1 on the input genes and 2) adding noise to gene expression values ($\varepsilon \sim \mathcal{N}(0,0.1)$).
The model was trained with a batch size of 64 for a maximum of 2000 epochs. 

\paragraph{SVAE.}
The model was trained on molecules provided in SMILES notation, the longest molecules had 1423 tokens.
Both encoder and decoder consisted of two layers of bidirectional GRU (hidden size of 128, dropout of 0.1 at the first layer), each complemented with 50 parallel memory stacks with the depth of 50.
The latent space of SVAE had the same dimensionality as the PVAE (128) to enable the addition of encodings.
Similar optimization parameters as PVAE were used.
This model further utilized teacher forcing~\cite{williams1989learning}, i.e., the model's output is conditioned on the previous ground truth sample as opposed to its generated output.
Whilst this significantly simplifies learning, it may drive the generator to predominantly rely on the decoder (thus neglecting the latent encoding).
This so called posterior collapse was resolved by applying a token dropout rate of 0.1 during teacher forcing as suggested by~\cite{bowman2015generating}.
In addition to token dropout, KL cost-annealing~\cite{bowman2015generating} was employed during training, 
The model was trained with a batch size of 128 for a maximum (early stopping) of $\sim$ 110,000 steps (i.e., exactly 10 epochs)
During training, KL cost-annealing as described in~\cite{bowman2015generating} was explored in order to trade-off reconstruction and KL loss.

\paragraph{Critic.}
The critic was trained using the parameters reported in~\cite{manica2019paccmann} and replicating the best performing architecture based on multiscale convolutional encoders.

\paragraph{RL training.}
In order to maximize Equation~\ref{eq:policy_gradient}, we employed Adam optimizer ($\beta_{1}=0.9$, $\beta_{2}=0.999$, $\varepsilon=1\mathrm{e}{-4}$, weight decay $1\mathrm{e}{-4}$) and a decreasing learning rate starting at $1\mathrm{e}{-5}$.
The gradients were clipped to 2 to prevent $\mathcal{G}$ from destroying its chemical knowledge about SMILES syntax obtained through pretraining on ChEMBL.
The reward function hyperparameter $\alpha$ was set to 5.

\subsection{Results for gene expression profile VAE and SMILES VAE}
\label{ssec:resultspretraining}
\begin{figure}[!htb]
\centering
    \includegraphics[width=1.\textwidth]{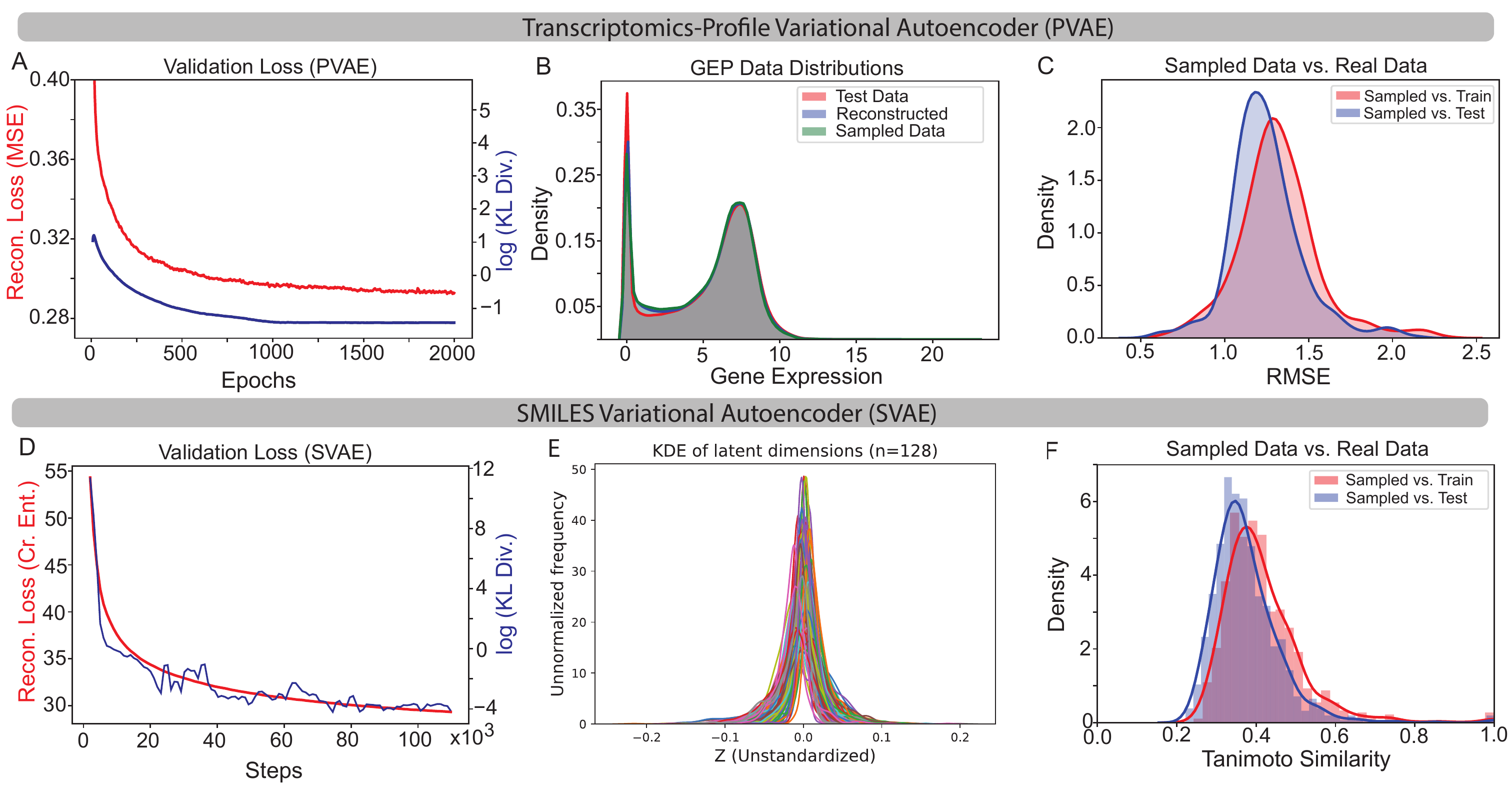}\\
    \caption{
    Results of pretrained PVAE and SVAE models.
    PVAE:
	(A) Development of validation error over the course of training.
    Reconstruction loss (MSE) and KL divergence are shown separately for comparison.
	(B) Distribution of gene expression values in real, reconstructed and generated samples.
	(C) Sampled (i.e. generated) data from the latent space of PVAE compared against training and test datasets from TCGA.
    SVAE:
	(D) Development of validation error over the course of training.
    Cross-entropy between target and generated SMILES is shown separately from the KL divergence (log scale for visual clarity).
    One epoch corresponds to $\sim$11 000 training steps.
	(E) Kernel density estimates of all 128 latent dimensions before decoding the test samples. 
	As enforced by the variational constraint, the latent variables follow Gaussian distributions.
	(F) The Tanimoto similarity between the Morgan fingerprints (ECFP) of the generated molecules and the structures from ChEMBL train and test datasets is used to verify that the generated compounds are sufficiently different from the training data.
    }
    \label{fig:vae_results}
\end{figure}

\paragraph{Profile VAE (PVAE)}
The pretraining results of the PVAE are presented in \autoref{fig:vae_results}A, B and C.
As shown in \autoref{fig:vae_results}B, the reconstructed gene expression profiles (GEP), shown in blue, as well as the generated GEPs (green) accurately mimic the distribution of the original GEPs (red).
Furthermore, the sampled GEPs follow the same lognormal distribution as the original data.
\autoref{fig:vae_results}C shows that the generated GEPs exhibit a higher similarity to the testing than to the training sample.
Overall, these results suggest that the PVAE learns to embed GEPs meaningfully into a latent space that allows both reconstruction and generation of new realistic GEPs of human cells. \\

\begin{figure}[!htb]
\centering
    \includegraphics[width=1\textwidth]{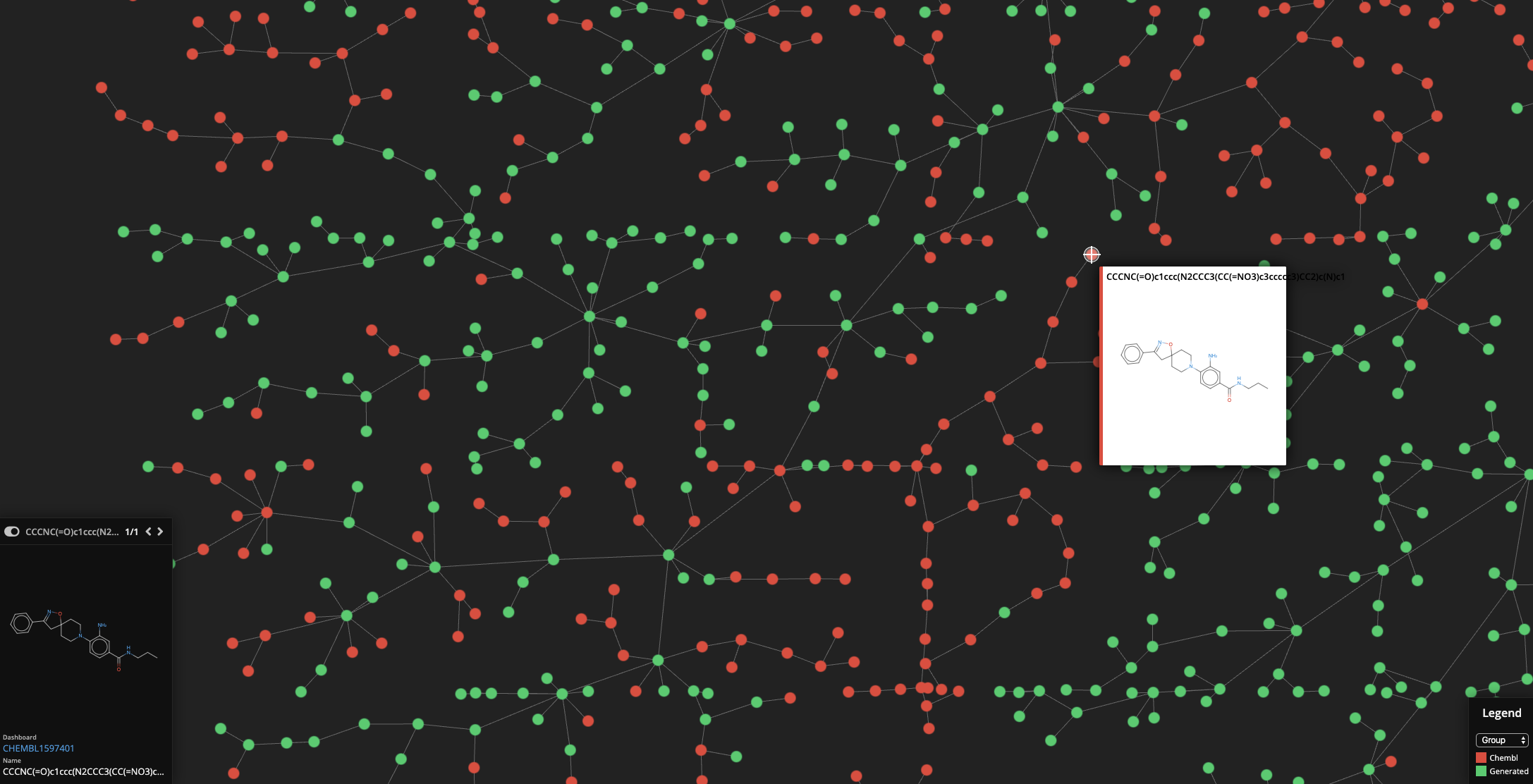}\\
    \caption{
    \textbf{Snapshot of the interactive TMAP visualization.}
    Generated molecules (green) and ChEMBL compounds (red) are shown through the TMAP algorithm which visualizes the chemical space by aggregating molecules with similar fingerprints (ECFP).
    The similarity in fingerprints is proportional to the distance of the repsective nodes on the spanning tree. 
    To explore the visualization interactively, please visit \href{https://paccmann.github.io/rl/unbiased}{https://paccmann.github.io/rl/unbiased}.
    }   
    \label{fig:tmap}
\end{figure}

\vspace{-2mm}
\paragraph{SMILES VAE (SVAE)}
\autoref{fig:vae_results}D, E and F give a quantitative analysis of the SVAE results following pretraining for 10 epochs with $\sim$1.4 million structures from ChEMBL.
To investigate the novelty and diversity of the generated molecules, we sampled 10,000 molecules by decoding random points from the latent space.
Overall, 96.2\% of the 10,000 generated molecules were valid molecular structures (assessed via \texttt{RDKit}) surpassing the results reported by~\citet{popova2018deep} who used the same stack-augmented GRUs trained on the ChEMBL database (95\% SMILES validity).
In addition, 99.72\% of the valid generated molecules were unique across the 10,000 generations.
The kernel density estimate (KDE) of the dimensions of the latent space (\autoref{fig:vae_results}E) validates that the SVAE fulfills the variational constraint as imposed by the Kullback-Leibler divergence in its loss function.
We then utilized a well-established chemical structure similarity measure, the Tanimoto similarity~\citep{Tanimoto1958elementary} to compare the ECFP~\citep{rogers_extended-connectivity_2010} of a subset of 1000 generated molecules with the training and test data from ChEMBL.
\autoref{fig:vae_results}F presents the distributions of the highest Tanimoto similarity between each generated compound and all compounds in training and test dataset respectively.
Only a negligible fraction of the generated molecules existed in either of the datasets, whereas the vast majority had a Tanimoto similarity ($\tau$) between 0.2 and 0.6 suggesting that our model learned to propose novel molecular structures from the chemical space of about 10\textsuperscript{30} to 10\textsuperscript{60} molecules~\citep{polishchuk2013estimation}. 
In addition, a \href{https://paccmann.github.io/rl/unbiased.html}{visualization of the chemical space} of ChEMBL as well as generated compounds through the TMAP algorithm (a library that visualizes high dimensional data through minimum spanning trees~\citep{probst2019visualization}) showed that the generated molecules mix well with the training molecules into the chemical space.
A snapshot of the interactive visualization is shown in~\autoref{fig:tmap}.

\begin{figure}[!htb]
\centering
    \includegraphics[width=1\textwidth]{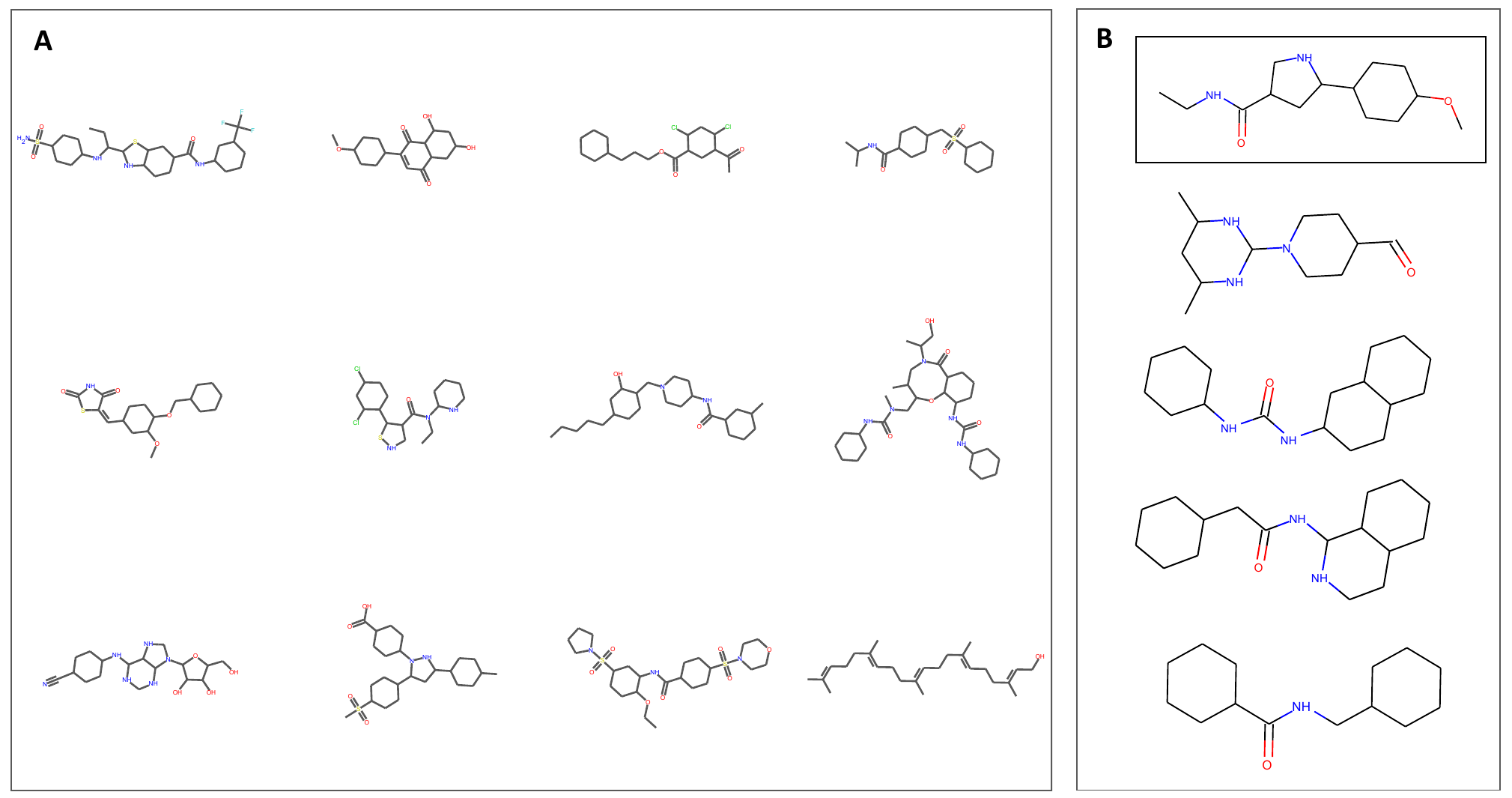}\\
    \caption{
    Qualitative inspection of generated molecules.
    (A) A sample of 12 molecular structures produced with the SVAE.
    (B) The molecule depicted at the top was encoded into the latent space.
    The four molecules below show different decodings from the latent space in the vicinity of the starting molecule.
    }   
    \label{fig:smiles_generator_qualitatitve}
\end{figure}

\autoref{fig:smiles_generator_qualitatitve}A, showcases a panel of 12 generated molecules for qualitative assessment of the molecular structures.
The generated molecules generally share drug-like structural features.
To inspect the smoothness of the latent space of molecules, we encoded a reference molecule shown at the top of \autoref{fig:smiles_generator_qualitatitve}B into the latent space and decoded four points in the vicinity of the reference molecule leading to the generation of structurally similar yet different compounds.

\FloatBarrier

\subsection{Nearest neighbors in ChEMBL}
\label{ssec:nnchembl}
\begin{figure*}[!htb]
    \includegraphics[width=1\textwidth]{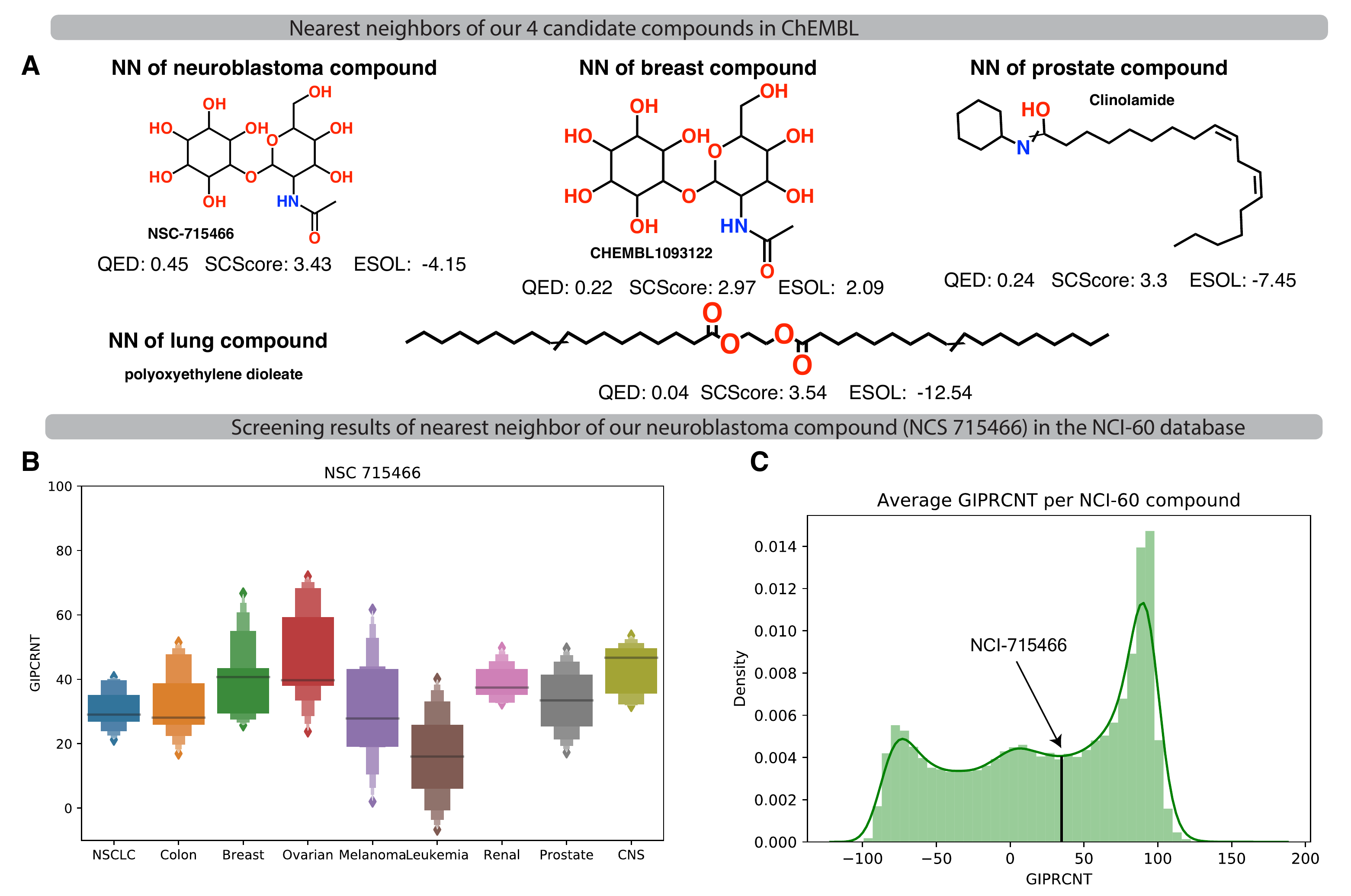}\\
    \caption{\textbf{A depiction of the nearest neighbors of our site-specific compounds in the ChEMBL database.}
    \textbf{A)} For all four site-specific generated compounds, we performed a similarity search across all molecules in the ChEMBL database. The nearest neighbors are depicted together with relevant drug-like properties.
    \textbf{B)} NSC-715466 was tested as part of the NCI-60 database~\citep{shoemaker2006nci60}, where it showed the strongest cell growth inhibition effect against leukemia cell lines.
    The GIPRCNT is a metric to measure cytotoxicity, where 100\% refers to unchanged cell proliferation (identical to the control cells), 0\% to complete stopping of cell proliferation and -100\% to a full inhibition of all cells.
    \textbf{C)} NSC-715466 showed only moderate anticancer effects as reported in the NCI-60 database.
    }
    \label{fig:chembl_panel}
\end{figure*}

To further validate the four site-specific compounds as proposed by our model, \autoref{fig:chembl_panel} depicts the respective nearest neighbors (measured by Tanimoto similarity) in the ChEMBL database of bioactive compounds.

\FloatBarrier
\subsection{Validation of critic (PaccMann) on ChEMBL data}
\label{ssec:paccmannchembl}
Our critic, PaccMann, is an anticancer drug sensitivity prediction model that has been trained \textit{only} on anticancer compounds from GDSC.
Since IC50 cell screening data for compounds with knowingly no anticancer effects are notoriously unavailable, PaccMann lacks a \textit{negative training set} which would help extending its generalization across the space of known anticancer compounds.
One could thus suspect that PaccMann is generally a flawed evaluator of compounds falling outside the space of anticancer drugs and that it would be biased towards predicting high efficacy for compounds without anticancer effects.

\begin{figure*}[!htb]
    \includegraphics[width=1\textwidth]{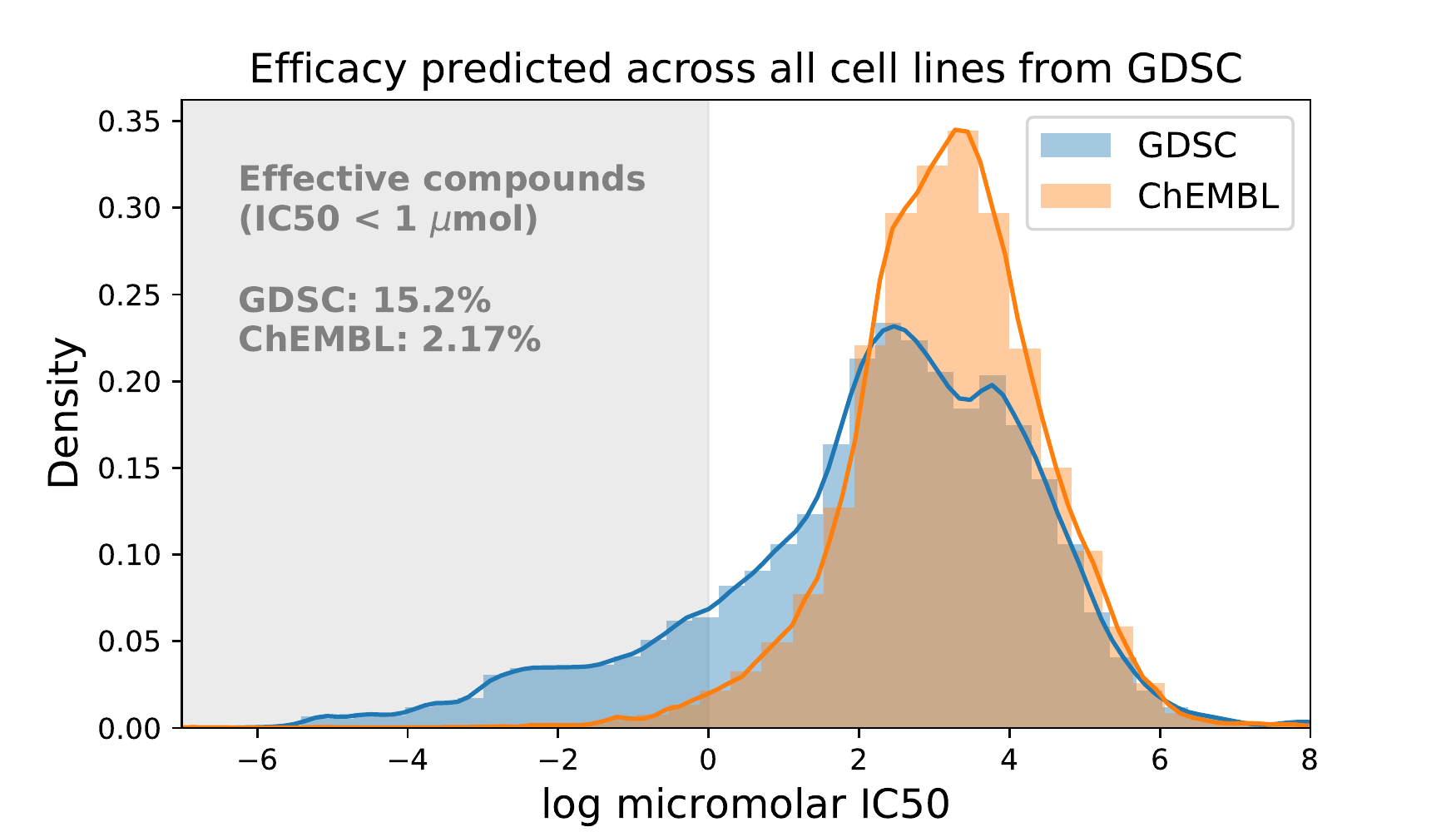}\\
    \caption{\textbf{A comparison of the predicted IC50 of GDSC drugs (known anticancer effects) and bioactive, drug-like compounds from ChEMBL. }
    The density plot of IC50 values as predicted by PaccMann from 210 cancer drugs from GDSC (blue) and 1000 molecules from ChEMBL (orange) across all 965 cell lines from the GDSC panel shows that only a minimal portion (2.17\%) of ChEMBL compounds are predicted as effective against a given cell line, whereas this holds for a significantly larger fraction of GDSC compounds.
    }
    \label{fig:gdsc_chembl}
\end{figure*}

For that reason, \autoref{fig:gdsc_chembl} shows the predicted efficacy of all anticancer compounds from GDSC (both training and testing data) as well as the predicted efficacy of a representative set of 1000 molecules from ChEMBL.
The predicted logarithmic IC50 across all cancer drugs was $2.2\pm2.2$ whereas it was only $3.2\pm1.6$ for ChEMBL molecules.

\FloatBarrier
\subsection{Possible synthesis routes for generated molecules}
\label{ssec:retrosynthesis}
\begin{figure*}
    \includegraphics[width=1.\textwidth]{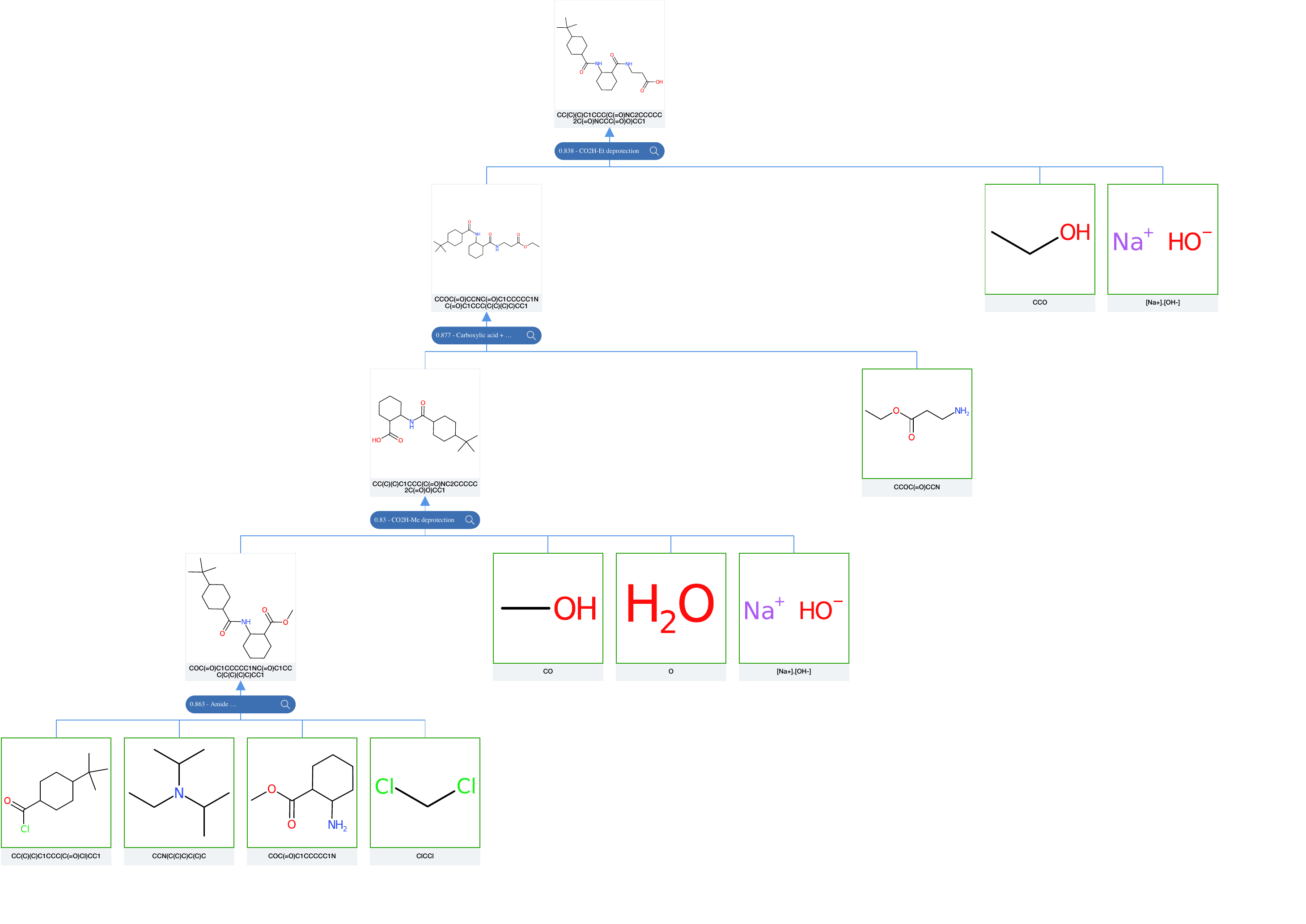}\\
    \caption{\textbf{A retrosynthesis route for a generated molecule.}
    A possible synthesis route, predicted by a molecular retrosynthesis model~\citep{ibmrxn} is shown for a compound proposed against nervous system cancer (top middle).
    The predicted synthesis consists of four sequential reactions with a total 10 commercially available reactants (green).
    }
    \label{fig:retrosynthesis}
\end{figure*}

In order to assess the complexity of a potential synthesis of compound proposed by our model \autoref{fig:retrosynthesis} shows a predicted synthesis route for a compound gen rated against nervous system cancer.
While the model proposes a panel of possible synthesis routes, the one depicted in~\autoref{fig:retrosynthesis} decomposes the synthesis into four reactions and a total of ten commercially available reactants.
The simplicity of the proposed route as well as the high confidence scores ($>0.8$) associated to each reaction seem to be promising indicators towards a possible synthesis.
Details of the synthesis route are depicted at the end of document.

\begin{figure*}
    \includegraphics[width=1.\textwidth]{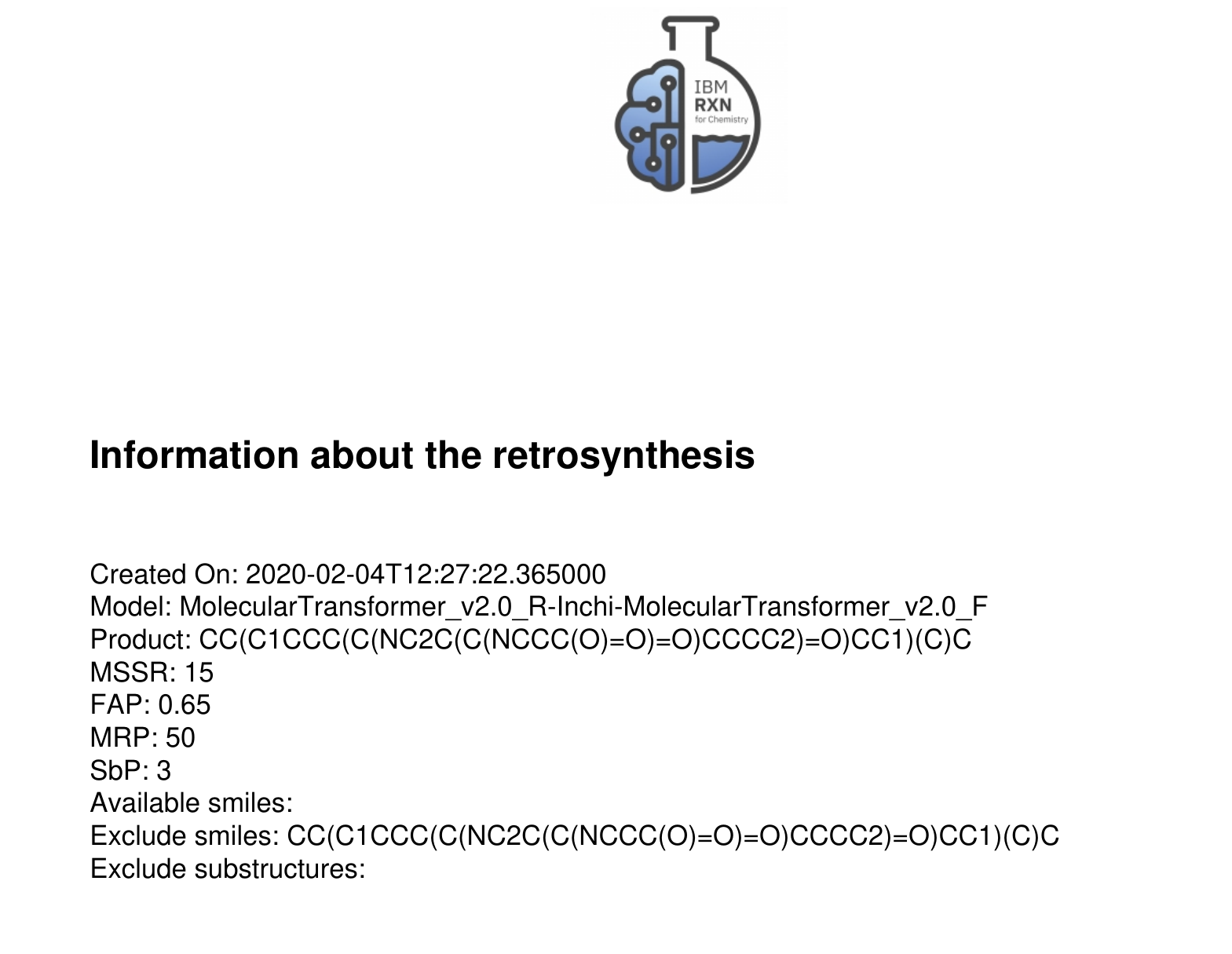}\\
    \label{fig:retro1}
\end{figure*}
\begin{figure*}
    \includegraphics[width=1.\textwidth]{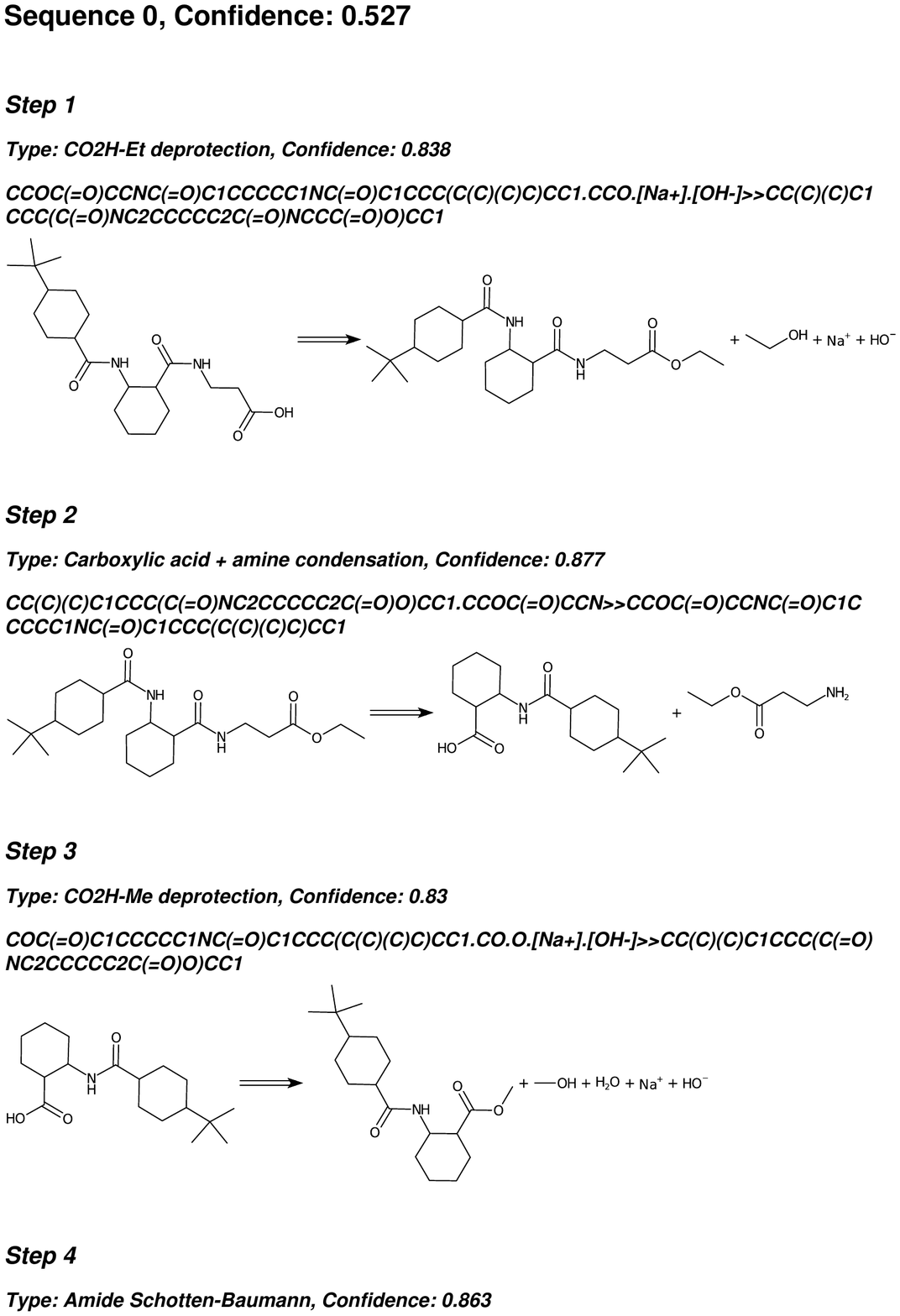}\\
    \label{fig:retro2}
\end{figure*}
\begin{figure*}
    \includegraphics[width=1.\textwidth]{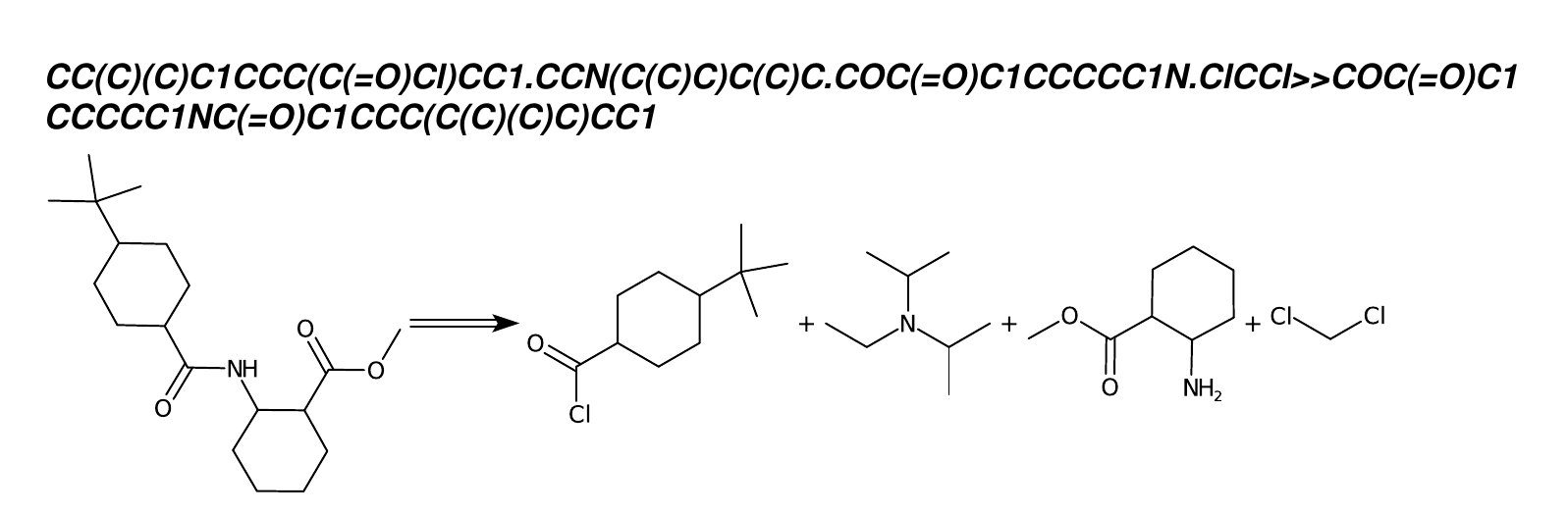}\\
    \label{fig:retro3}
\end{figure*}
\thispagestyle{plain}

\end{document}